\documentclass[11pt]{elsarticle}

\usepackage{amssymb}
\usepackage{amsmath}
\usepackage{braket}
\usepackage{ulem}

\usepackage{subcaption}
\usepackage[inline]{enumitem}
\usepackage{booktabs}
\usepackage{multirow}
\usepackage[version=4]{mhchem}
\usepackage{rotating}
\usepackage{tabularx}
\usepackage{makecell}
\usepackage{array}

\usepackage{algorithm}
\usepackage{algorithmicx}
\usepackage{algpseudocode}
\usepackage{mdframed}
\usepackage{bm}

\floatstyle{plain}
\restylefloat{algorithm}

\usepackage{caption}
\makeatletter
\newlength{\InnerAlgorithmCaptionAbove}
\newlength{\InnerAlgorithmCaptionBelow}
\makeatletter
\newcommand{\InnerAlgorithmCaption}[2]{%
  \refstepcounter{algorithm}%
  \label{#2}%
  \addcontentsline{loa}{algorithm}{%
    \protect\numberline{\thealgorithm}{#1}}%
  \par\vspace{\InnerAlgorithmCaptionAbove}%
  \@fs@capt{\fnum@algorithm}{#1}%
  \par\vspace*{\InnerAlgorithmCaptionBelow}%
}
\makeatother

\usepackage{hyperref}
\usepackage{xcolor}
\usepackage{xspace}
\usepackage{microtype}
\usepackage[margin=3.5cm]{geometry}

\newcommand{\abinit}{\textsc{Abinit}\xspace}

\newcommand{\blas}{BLAS\xspace}
\newcommand{\lapack}{LAPACK\xspace}
\newcommand{\openmp}{OpenMP\xspace}
\newcommand{\openacc}{OpenACC\xspace}
\newcommand{\gemm}{\texttt{gemm}\xspace}
\newcommand{\hegvd}{\texttt{hegvd}\xspace}
\newcommand{\potrf}{\texttt{potrf}\xspace}
\newcommand{\trsm}{\texttt{trsm}\xspace}
\newcommand{\nvidia}{NVIDIA\xspace}
\newcommand{\amd}{AMD\xspace}
\newcommand{\openmptarget}{OpenMP \texttt{target}\xspace}
\newcommand{\openmpoffload}{OpenMP offloading\xspace}
\newcommand{\mpi}{MPI\xspace}
\newcommand{\hip}{HIP\xspace}
\newcommand{\cublas}{cuBLAS\xspace}
\newcommand{\cufft}{cuFFT\xspace}
\newcommand{\rocblas}{rocBLAS\xspace}
\newcommand{\rocfft}{rocFFT\xspace}
\newcommand{\cusolver}{cuSOLVER\xspace}
\newcommand{\rocsolver}{rocSOLVER\xspace}
\newcommand{\rocm}{ROCm\xspace}
\newcommand{\cuda}{CUDA\xspace}
\newcommand{\cray}{Cray\xspace}

\newcommand{\nvhpc}{NVHPC\xspace}
\newcommand{\topaze}{\textit{Topaze}\xspace}
\newcommand{\jeanzay}{\textit{Jean Zay}\xspace}
\newcommand{\adastra}{\textit{Adastra}\xspace}
\newcommand{\lobpcg}{LOBPCG\xspace}
\newcommand{\chebyshevfiltering}{Chebyshev filtering\xspace}
\newcommand{\alltoall}{all-to-all communication\xspace}
\newcommand{\kpoint}{$\bm{\mathsf{k}}$-point\xspace}
\newcommand{\sdeflate}{\texttt{S-Deflate}\xspace}
\newcommand{\sortho}{\texttt{S-Ortho}\xspace}
\newcommand{\rr}{\texttt{RR}\xspace}
\newcommand{\mpia}{{\color{darkpink}\textbf{mpi-A}}\xspace}
\newcommand{\mpir}{{\color{cfblue}\textbf{mpi-R}}\xspace}
\newcommand{\mpisr}{{\color{darkgreen}\textbf{mpi-sR}}\xspace}
\newcommand{\ttH}{\texttt{H}}
\newcommand{\ttS}{\texttt{S}}

\definecolor{cfblue}{HTML}{005A9C}
\definecolor{darkpink}{HTML}{9C005A}
\definecolor{darkgreen}{HTML}{5A9C00}
\definecolor{capri}{rgb}{0.0, 0.75, 1.0}

\journal{Computer Physics Communications}

\begin{document}

\begin{frontmatter}

\title{GPU acceleration of plane-wave density functional theory calculations in \abinit}

\author[label1,label2]{Ioanna-Maria Lygatsika\corref{cor1}\fnref{footn}}
\ead{ioanna-maria.lygatsika@inria.fr}
\author[label3]{Marc Sarraute}
\ead{marc.sarraute@asplus.fr}
\author[label1,label2]{Lucas Baguet}
\ead{lucas.baguet@cea.fr}
\author[label1,label4]{Pierre Kestener}
\ead{pierre.kestener@cea.fr}
\author[label1,label2]{Marc Torrent}
\ead{marc.torrent@cea.fr}

\fntext[footn]{Present address: COMMEDIA, Laboratoire Jacques-Louis Lions, Sorbonne Université and Inria Paris, 48 rue Barrault, 75013 Paris, France.}
\cortext[cor1]{Corresponding author.}

\affiliation[label1]{organization={CEA DAM-DIF},
             city={Arpajon},
             postcode={F-91297},
             country={France}}

\affiliation[label2]{organization={Université Paris-Saclay, CEA, LMCE},
             city={Bruyères-le-Châtel},
             postcode={F-91680},
             country={France}}

\affiliation[label3]{organization={Alliance Services Plus},
             addressline={62 rue Emile Zola},
             city={Boulogne-Billancourt},
             postcode={F-91200},
             country={France}}

\affiliation[label4]{organization={Université Paris-Saclay, CEA, LiHPC},
             city={Bruyères-le-Châtel},
             postcode={F-91680},
             country={France}}

\begin{abstract}
We report on the GPU port of the \abinit high-performance simulation code for plane-wave DFT calculations. Large-scale electronic structure calculations require computing the electronic wave function by solving the Kohn-Sham equations discretized over a large number of plane waves. Porting such calculations to GPU nodes relies not only on extensive usage of vendor libraries from a development perspective, but also on algorithmic revisions of the iterative diagonalization procedure in the resolution of the Kohn-Sham equations to identify GPU-efficient mathematical operations (linear algebra, FFTs) applied to the wave function distributed in memory. The present contribution discusses the \abinit implementation on multi-GPU architectures, providing detailed performance results for heterogeneous CPU-GPU nodes versus CPU nodes. Particular attention is given to comparing two diagonalization algorithms---Locally Optimal Block Preconditioned Conjugate Gradient and Chebyshev polynomial filtering---in terms of GPU efficiency.
\end{abstract}


\end{frontmatter}

\section{Introduction}

Large-scale simulations in materials science at the quantum level involve the computation of thousands of electronic states. This raises a challenge when it comes to efficiently exploiting the latest High-Performance Computing (HPC) architectures for reducing the simulation time, mainly due to the large problem size associated with the number of states and basis functions. Over the past decade, graphics processing units (GPU) have become widely available as accelerators in modern HPC systems. The advent of accelerators comes on top of an existing parallel programming landscape relying on multi-core architectures. The progressive evolution of hardware has led HPC simulation code developers to adopt new hybrid programming models for heterogeneous architectures. As GPU vendor libraries have become increasingly available, the porting task is facilitated to a great extent. Electronic structure calculations using \textit{ab initio} methods, namely Kohn-Sham Density Functional Theory (DFT) \cite{PhysRev.136.B864,PhysRev.140.A1133}, are a great candidate for GPU architectures due to the computationally expensive linear algebra operations required to calculate the electronic wave function, specifically the diagonalization of Hamiltonians in a large basis set, namely plane waves, which GPUs accelerate efficiently.

Electronic-structure DFT codes are, however, not all of the same nature. They differ according to the physical approximation, the basis, and the discretization space used to represent the Kohn-Sham equations. Some codes are designed for periodic solids, whereas others target molecular systems or quantum chemistry. Some operate in reciprocal space with plane waves, while others use real-space grids, finite differences, finite elements, wavelets, or localized atomic orbitals. Some treat all electrons explicitly, while others rely on pseudopotentials to replace core electrons with an effective potential. These differences strongly affect the GPU porting strategy, because they determine the dominant kernels, the data structures, the communication pattern, and the nature of the eigensolvers. Recent reviews provide a broader perspective on GPU acceleration in electronic structure and related numerical methods \cite{Fattebert2024,LEVITT201598}.

In recent years, several \textit{ab initio} codes have been ported to GPU, including plane-wave codes such as \textsc{Quantum ESPRESSO} \cite{QE_GPU,QE_EXA}, \textsc{VASP} \cite{VASP2012,VASP2018,VASP2024}, and \textsc{ABINIT} \cite{Verstraete2025}; real-space codes such as \textsc{GPAW} \cite{GPAW2024}; wavelet-based codes such as \textsc{BigDFT} \cite{genovese2011improvements}; finite-element approaches \cite{DAS2019,DAS2022}; and all-electron full-potential codes such as \textsc{Exciting} \cite{exciting_gpu_2024}. Although these codes all address the Kohn-Sham DFT equations, their GPU porting challenges are distinct. In plane-wave DFT, the Hamiltonian is applied in a matrix-free fashion and the Hamiltonian matrix is never formed explicitly because it is dense and cannot be stored explicitly. In real-space and finite-element formulations, the sparsity pattern and the locality of the discretization lead to different parallel kernels and different solver bottlenecks. In all-electron codes, the treatment of the core region and the basis representation impose additional constraints not present in pseudopotential plane-wave codes.

This paper focuses on \abinit, a widely used open-source simulation package for materials science based on plane-wave \textit{ab initio} methods \cite{abinit_web,Verstraete2025}. ABINIT has a long history of adaptation to modern supercomputer architectures, using \mpi, hybrid \mpi-\openmp parallelism on CPUs, and more recently CPU-GPU hybridization \cite{GAVINI2023}. The GPU porting effort discussed here belongs to a broader modernization strategy and aims at providing a portable implementation that can target different accelerator vendors through a uniform programming model.

Several programming-model strategies exist for porting a CPU code to GPU accelerators, each with different trade-offs between performance, portability and development effort. Direct, vendor-specific implementations, such as \cuda for \nvidia GPUs~\cite{nvidia_cuda} or \hip~\cite{rocm_hip} for \amd GPUs, generally give the best performance on a given architecture but require dedicated kernels and offer limited portability across vendors. Directive-based approaches, such as \openmptarget offloading~\cite{openmp_spec} and \openacc~\cite{openacc_spec}, instead annotate existing CPU source code with compiler pragmas, lowering the development cost and preserving a single source code, at the possible expense of some performance compared to hand-written kernels. Performance-portability libraries, such as Kokkos~\cite{CARTEREDWARDS20143202}, RAJA~\cite{Beckingsale2019} or SYCL~\cite{sycl2020}, offer a further alternative: they expose C++ abstractions for parallel loops and data structures that are compiled down to the appropriate backend at build time, at the cost of adopting a dedicated programming ecosystem.

The present work focuses on \openmp offloading implementation of plane-wave DFT, as done in the \abinit software, motivated by portability, incremental adoption, and the availability of mature compiler support on \nvidia and \amd platforms. Other plane-wave DFT codes have also used \openmp-based GPU offloading, notably \textsc{Quantum ESPRESSO} \cite{QE_GPU,QE_EXA}. However, in most reported \openmp GPU ports, the effort has concentrated primarily on the offloading of SIMD (Single Instruction, Multiple Data) operations such as FFTs and dense linear algebra kernels to the GPU rather than on revisiting the numerical algorithms used at the core of the self-consistent field cycle. We argue that there remains room to go further in this direction, in particular by examining how the choice of iterative eigensolver interacts with GPU porting in plane-wave DFT, where, as noted above, the Hamiltonian is never assembled and only matrix-free Hamiltonian-vector products are available.

Iterative eigensolvers for Kohn-Sham DFT can be broadly classified into minimization methods and diagonalization algorithms. The former include conjugate-gradient, Davidson, \lobpcg{}~\cite{knyazev2001} and RMM-DIIS-like schemes. The latter include subspace and filtered methods based on spectral projectors or polynomial filters, such as FEAST \cite{polizzi2009}, \chebyshevfiltering{}~\cite{zhou2006}, and spectrum slicing methods~\cite{SCHOFIELD2012497}.
For plane-wave DFT specifically, Levitt and Torrent~\cite{LEVITT201598} showed how a \chebyshevfiltering-based eigensolver can outperform block conjugate-gradient methods and scale to tens of thousands of CPU cores, precisely because it reduces reliance on global orthogonalization between blocks of bands,a property which is also decisive for exposing the fine-grained parallelism GPUs require. Some of these eigensolvers have already been ported to GPU, notably the Chebyshev-filtered subspace iteration used in the finite-element code \textsc{DFT-FE}~\cite{DAS2022}; to our knowledge, however, no similar GPU study directly comparing LOBPCG and a Chebyshev filtering eigensolver within the same plane-wave DFT code has previously been reported.

Parallelization also depends on the discretization and on the physical model. In plane-wave DFT, beyond the obvious parallelism over \kpoint and spin channels, the main sources of concurrency are: \begin{enumerate*}[label=(\roman*)] \item the intrinsic parallel scalability of the mathematical libraries involved, essentially FFT and dense linear algebra, and \item the ability to apply the Hamiltonian to several wave functions concurrently, which depends directly on how far the chosen iterative solver can decouple the treatment of individual wave functions from one another, and hence on how much global orthogonalization it requires at each step. \end{enumerate*} In real-space, finite-difference, and finite-element codes, similar opportunities exist, but additional locality-driven parallelism and different communication patterns may dominate \cite{GPAW2024,DAS2019,Fattebert2024}. As a result, the optimal GPU strategy is not universal across electronic-structure codes.

Plane-wave pseudopotential codes replace the all-electron Hamiltonian by a smoother pseudo-Hamiltonian acting on pseudo-wave functions, in order to keep the plane-wave basis size tractable. Different pseudopotential formalisms coexist: norm-conserving pseudopotentials~\cite{HSC1979}, ultrasoft pseudopotentials~\cite{Vanderbilt1990}; and the projector augmented-wave (PAW) method~\cite{Blochl1994}. Although these formalisms rest on slightly distinct theoretical constructions, they share a common computational structure and have comparatively little impact on the GPU porting strategy itself. Ultrasoft pseudopotentials and the PAW method do share one specific feature: because their pseudo-wave functions are not orthogonal, they lead to a generalized eigenvalue problem involving an overlap operator, which generally must be inverted to recast the problem into a standard eigenvalue form~\cite{LEVITT201598}. In practice, however, this additional step has only a marginal impact on overall performance.

The first GPU version of \abinit (v6.12) was implemented in 2012 ; the code was tested and successfully ported to \nvidia Fermi accelerators. This early GPU port was based on custom \cuda kernels, with batched FFTs and iterative diagonalization algorithms using the MAGMA library~\cite{magma} for \nvidia only. In this paper, we present the current GPU implementation of \abinit (v10.8), which has been re-implemented from scratch over the last two years using \openmp offloading and vendor libraries for the most performance-critical operations, focusing on the algorithmic and programming choices that make the port effective on modern accelerators.

\smallskip

We summarize our contributions as follows:
\begin{itemize}
    \item \textit{Overview of porting strategy.} We present porting concepts focusing on specific parts of the code successfully ported to GPU: matrix-matrix multiplications, FFTs, linear algebra operations.
    \item \textit{GPU-resident wave function calculations.} The wave function is updated regularly and maintained in GPU memory. We explain how we optimized memory usage and exploited batch processing for this calculation. We provide \mpi communication theoretical estimates for two iterative diagonalization algorithms: Locally Optimal Block Preconditioned Conjugate Gradient (\lobpcg) and Chebyshev polynomial filtering.
    \item \textit{GPU performance metrics.} We introduce metrics that allow to examine wave function calculation steps under the prism of arithmetic intensity and GPU memory bandwidth. We report an in-depth comparison of diagonalization algorithms, focusing on execution time, energy consumption and roofline models on GPUs, for assessing performance portability and comparing energy efficiency between vendors.
\end{itemize}

\section{Porting framework}\label{sec:gen_port}

This section presents the key considerations and computational design patterns at the core of \abinit{}'s GPU porting, irrespective of the programming model. We highlight the computational structure underlying wave function calculations and identify the shared characteristics of numerically intensive workloads that enabled efficient GPU porting, making our application well suited for GPU architectures. Additionally, we describe data movement concepts employed in GPU memory.

Firstly, Section~\ref{sec:batch} focuses on batch processing as key computational principle in \abinit. Section~\ref{sec:transfer} describes host-device data transfers. Section~\ref{sec:MPI_between_GPU} details the data distribution across GPUs and the MPI communication pattern between them. Lastly, Section~\ref{sec:limdims} briefly outlines the problem sizes dimensioning \textit{compute-bound} and \textit{memory-bound} operations and kernels, such as large matrix operations, prior to examining them in detail in Section~\ref{sec:iterative}.

\subsection{Batch processing} \label{sec:batch}

Batch processing programming approach is crucial for the GPU port, as it determines how routines and kernels execute when computing the primary object of interest---the electronic wave function $\Psi$. Numerical solvers make use of batch processing at the level of plane waves and of electronic bands, i.e., rows and columns of $\Psi$ respectively. Throughout this document, let $N$ denote the number of plane waves and $M$ the number of electronic bands $\psi_i$.

In essence, batch processing consists of regrouping data to operate on larger chunks instead of many smaller ones. Regrouping the data exposes coarser-grained data parallelism, so as to provide a workload better suited to exploiting the computational power of GPUs. A requirement is that the task applied to each batch must be independent for all data and able to run asynchronously, typically in the sense of the SIMD (Single Instruction, Multiple Data) programming model. The use of batch processing is crucial for limiting device memory (DRAM) traffic overhead (bytes transferred) and for improving GPU kernel throughput.

The most representative example of task on batched data is the application of the plane-wave Hamiltonian operator, that can be executed on multiple electronic bands at once instead of per band in a loop. The application of the local part of the plane-wave Hamiltonian, denoted by $V_{\rm loc}$, to the wave function uses the fast Fourier transform (FFT). The main idea is to execute many independent FFTs asynchronously instead of relying on parallel FFT implementations. To do this, we treat the matrix $\Psi$ as a batch of column vectors $\psi_i$ and apply FFT to each column independently by a single kernel call. Porting this operation on GPU follows the batched execution illustrated in Algorithm~\ref{algo:batched_fft}, compared to a less efficient band-wise execution shown in Algorithm~\ref{algo:perband_fft}. Low-level computational routines rely on FFTW library \cite{frigo1998fftw} on CPUs, cuFFT \cite{cufft} on NVIDIA GPUs, and rocFFT \cite{rocfft} on AMD GPUs (Section~\ref{sec:lowlevelops}).

\begin{algorithm}[H]
\centering
\setlength{\arrayrulewidth}{0.3pt}         
\setlength{\tabcolsep}{4pt}                
\footnotesize

\begin{tabular}{|c|}
\hline

\begin{minipage}[t]{0.65\linewidth}
    \begin{algorithmic}[1]
        \Require{$M$ electronic bands $\psi_i$}
        \Ensure{$V_{\rm loc}\psi_1,\ldots,V_{\rm loc}\psi_M$}
        \For{$\psi_i\in\psi_1,\ldots,\psi_M$}
            \State $\psi_i \gets \text{ApplyLocalPotential}(\psi_i)$\hfill\texttt{fftw\_plan\_dft\_3d} (\textit{CPU})
            \Statex \hfill \texttt{[cu/roc]fftPlan3D} (\textit{GPU})
        \EndFor
    \end{algorithmic}
    \footnotesize
    \InnerAlgorithmCaption{Per-band kernel execution.}{algo:perband_fft}
\end{minipage}

\\ \hline

\begin{minipage}[t]{0.65\linewidth}
    \begin{algorithmic}[1]
        \Require{electronic wave function $\Psi$}
        \Ensure{$V_{\rm loc}\Psi$}
        \State $\Psi\gets\text{ApplyLocalPotential}(\Psi)$\hfill\texttt{fftw\_plan\_many\_dft} (\textit{CPU})
        \Statex \hfill \texttt{[cu/roc]fftPlanMany} (\textit{GPU})
    \end{algorithmic}
    \InnerAlgorithmCaption{Batched bands kernel execution.}{algo:batched_fft}
\end{minipage}
\\ \hline
\end{tabular}
\end{algorithm}

The \abinit code previously implemented batch processing on CPUs for vectorized architectures within a single core~\cite{GONZE2016106, GONZE2020107042}. This optimization consisted in regrouping data into batches, that allowed to replace level-2 \blas operations by level-3 ones and FFTs to multiple vectors of plane-wave coefficients at once. The wave function is thus globally treated by blocks on CPU and the GPU port was greatly facilitated by this existing feature, especially adapted to massively parallel architectures.

While level-3 \blas and batched 3D FFTs for treating electronic bands are optimizations also found in the CPU implementation, we mention the following new batching design particularly well suited to GPU-resident algorithms. The GPU port of \abinit goes a step further by making use of many batched kernel calls on smaller matrices \cite{dongarra_batch}. This new design is adapted to highly parallel workloads as data is stored in a configurable stride in memory, where the stride is the distance between two consecutive electronic bands. An example of mathematical operation enabling this strided batch processing is batched matrix-matrix multiplication when applied to a group of small matrices independently. The latter is a new operation only present on the GPU port and for this reason it is discussed in more detail in Section~\ref{sec:lowlevelops}.

\subsection{Host-device memory transfers} \label{sec:transfer}

The heaviest data move between host and device happens at the wave-function level. As a consequence, it is critical to offload most wave-function-based computations to the GPU while minimizing host-device data transfers. In order to port the wave function calculation to GPU, the wave function must first be transferred to GPU. The main strategy consists in performing a single transfer of the wave function from CPU to GPU at the start of every self-consistent field (SCF) iteration for the current \kpoint of the reciprocal space. Then we avoid any intermediate transfers and keep the wave function for that \kpoint entirely on GPU memory to perform the diagonalization of a constant Hamiltonian operator. Algorithm~\ref{algo:scf} shows the host-device memory transfers (H2D denoting host-to-device and D2H device-to-host) in main wave-function computation for GPU-resident diagonalization (step 5) performed iteratively within SCF outer loop. $H$ is the plane-wave Hamiltonian at current SCF iteration and $S$ the PAW overlap matrix \cite{Blochl1994}.

\begin{algorithm}[H]
\centering
\footnotesize
\setlength{\arrayrulewidth}{0.3pt} 
\setlength{\tabcolsep}{4pt}        

\begin{tabular}{|c|}
\hline
\begin{minipage}[t]{0.9\linewidth}
    \begin{algorithmic}[1]
        \State Initialize $\Psi$. \hfill\textit{(on CPU)}
        \While{self-consistent field iteration}
        \For{every \kpoint{}}
        \State {\color{cfblue}\textbf{Copy}} $\Psi$ for current \kpoint from CPU to GPU.\hfill\textit{(H2D)}
        \State Solve $H\Psi=\Lambda S\Psi$ for current \kpoint.\hfill\textit{(solver on GPU)}
        \State {\color{cfblue}\textbf{Copy}} $\Lambda$ for current \kpoint from GPU to CPU.\hfill\textit{(D2H)}
        \EndFor
        \State Compute electronic occupations $f$ from $\Lambda$.\hfill\textit{(on CPU)}
        \For{every \kpoint{}}
        \State Accumulate electronic density from $f,\Psi$ for current \kpoint.\hfill\textit{(FFT on GPU)}
        \EndFor
        \State \textbf{if} convergence met \textbf{then} exit
        \EndWhile
        \State \textbf{return} $\Psi$
    \end{algorithmic}
    \InnerAlgorithmCaption{Self-consistent field in DFT.}{algo:scf}
\end{minipage}
\\ \hline
\end{tabular}
\end{algorithm}

Once all \kpoint{}s have been treated, the electronic occupations are computed from the eigenvalues (step 8 of Algorithm~\ref{algo:scf}). This step is performed on the CPU, since its computational cost is negligible. For metallic systems, it requires gathering the eigenvalues from all GPUs, as computing the occupations requires the full set of eigenvalues; this represents a comparatively small host-device transfer, and no wave-function data is involved. The electronic density is then computed from the occupations and the converged wave function coefficients for every \kpoint{}, using an FFT operation on GPU. In most cases, the wave functions are still resident on the GPUs that computed them, so this step requires no additional host-device transfer; only in extreme memory-constrained cases might some wave functions need to be reloaded onto the GPU at this stage.

\subsection{Communication between GPUs}\label{sec:MPI_between_GPU}

In the GPU port of \abinit, all \mpi data transfers are implemented using a GPU-aware \mpi implementation, i.e. allowing hardware-specific optimizations, and support GPU pointers. To every \kpoint, we associate a set of parallel processes. The \mpi distribution of the wave function for a given \kpoint is implemented in a 2D processor grid as follows. The matrix of column vectors---each representing the coefficients of a wave function in the plane-wave basis---is uniformly distributed across a user-defined number $p$ of \mpi processes. In \abinit, each \mpi task is associated with a single GPU (note that several \mpi tasks can share the same GPU). Two types of wave-function distributions are required, depending on the computation:
\begin{itemize}
    \item Row-distributed, suited to operations involving all columns, or electronic bands, such as linear algebra solvers, and
    \item Column-distributed, suited to the application of the Hamiltonian and FFTs since each process holds the full set of plane waves.
\end{itemize}

\begin{figure}[H]
    \centering
    \includegraphics[width=\textwidth]{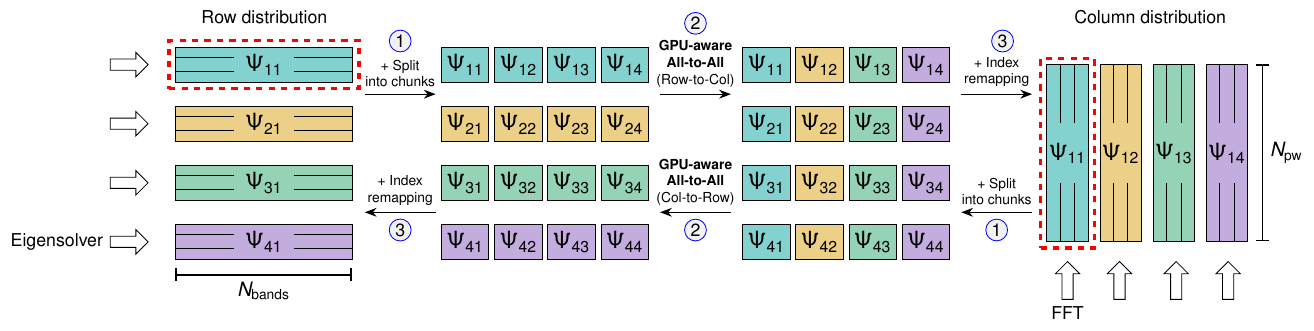}
    \caption{Row-to-column and column-to-row \mpi transpositions of the wave function $\Psi$ distributed across $p=4$ processes. Data transfer is managed by GPU-aware MPI library. Mathematical operations (FFTs, eigensolver) are executed per block (red dashed border) in parallel. In practice, we perform all-to-all-v communication (\mpia) to distribute plane-wave indices as uniformly as possible during row splitting.}
    \label{fig:mem_wf}
\end{figure}

Switching between wave-function \mpi distributions is executed by performing an all-to-all (\mpia) collective communication \cite{BOTTIN2008329}. The default distribution is by rows. Figure~\ref{fig:mem_wf} illustrates the \mpi transposition operation. Host-device memory transfers are executed only before and after all-to-all communications, thus the two can be entirely decoupled when GPU-aware \mpi is enabled.

\subsection{Limiting dimensions} \label{sec:limdims}

Limiting dimensions are responsible for bottlenecks and condition scaling. These dimensions, along with mathematical operations most highly affected, are:
\begin{itemize}
    \item number of electronic bands: linear algebra operations used for orthogonalization and diagonalization procedures,
    \item number of plane waves: size of FFTs (assuming batched band execution), matrix-matrix multiplications.
\end{itemize}

Note that \kpoint treatment is highly efficient and requires minimal communication, and is therefore not a limiting dimension. Fortunately, the mathematical operations constrained by limiting dimensions are inherently GPU-friendly, suggesting strong portability potential for our code. The following section further explains how limiting dimensions affect algorithms.

\section{Iterative algorithms}\label{sec:iterative}

Iterative diagonalization is key in plane-wave DFT to compute electronic wave functions using so-called \textit{matrix-free} algorithms. The solvers for this step are the most time-consuming part of plane-wave based electronic structure calculations. In this section, we present the parallel design of available solvers in \abinit and provide theoretical estimates to model their cost, in terms of computation in FLOPS (floating point operations per second), \mpi communications (message count and communication volume) and arithmetic intensity. We then use these theoretical estimates to predict the GPU-efficient parts and bottlenecks of our solvers from an algorithmic perspective.

\subsection{Problem setup}\label{sec:problemsetup}

The Kohn-Sham equations for electronic wave functions $\psi_i$ ($i=1,\ldots,M$) read $H\psi_i = \lambda_i S\psi_i$, where $H$ is the Hamiltonian operator and $S$ the overlap operator for the $\psi_i$. $S = I$ only with norm-conserving pseudopotentials; $S \neq I$ whenever the norm constraint is relaxed, as in Projector Augmented-Wave (PAW) approach \cite{Blochl1994}. $H$ and $S$ are $N\times N$ Hermitian matrices and $\Psi=(\psi_i)_{1\leq i\leq M}$ is a $N\times M$ matrix of wave functions expressed in plane wave basis. We assume in this section that the electronic density is fixed and we are within a fixed SCF iteration, that is, in step 4 of Algorithm~\ref{algo:scf}. We look for the first smallest $M\ll N$ eigenpairs of $H$. The iterations used within the diagonalization will be referred to as \textit{inner} iterations to distinguish from SCF \textit{outer} iterations, not discussed here.

\subsection{Main computational kernels}

We describe here the core components of available iterative solvers in \abinit from a mathematical point of view, independently of their implementation.

\paragraph{Eigensolvers} All iterative diagonalization algorithms for the Hamiltonian follow a two-step process:

\begin{enumerate}[label=(\roman*)]
\item Subspace iteration: Build a subspace by iterating on trial vector blocks, e.g., via Krylov methods, conjugate gradient (CG) residual minimization, or spectral filtering.
\item Eigenvector extraction: Diagonalize the Hamiltonian inside the subspace spanned by trial vectors to extract an approximate eigenspace using the \textit{Rayleigh-Ritz} (RR) procedure.
\end{enumerate}

The first step iteratively refines a set of trial vectors to approximate an invariant subspace, while the second step is a direct solver that yields the directions inside the invariant subspace that best approximate the actual eigenbasis. The Rayleigh-Ritz procedure for solving the generalized eigenvalue problem $H\Psi=\Lambda S\Psi$, where $\Lambda$ is a diagonal matrix with Ritz values in its diagonal, is illustrated in Algorithm~\ref{fig:algo_RR}. It consists of first projecting the Hamiltonian to the trial subspace, then solving a smaller generalized eigenproblem on the subspace, finally rotating the obtained solution to recover approximated eigenvectors.

\begin{algorithm}[H]
  \centering
  \footnotesize
  \setlength{\arrayrulewidth}{0.3pt} 
  \setlength{\tabcolsep}{4pt}        
  \begin{tabular}{|c|c|}
  \hline
  \begin{minipage}[t]{0.38\linewidth}
    \begin{algorithmic}[1]
        \Require{Matrices $H,S$, trial vectors $\Psi$.}
        \Ensure{$\Lambda,\Psi$.}
        \State $H_{\Psi} =\Psi^\top H\Psi$
        \State $S_{\Psi} = \Psi^\top S\Psi$
        \State $H_{\Psi}X = \Lambda S_{\Psi}X$ \hfill \hegvd
        \State $\Psi\gets X^\top \Psi$
        \vspace{8.3ex}
    \end{algorithmic}
    \footnotesize
    \InnerAlgorithmCaption{Rayleigh-Ritz procedure.}{fig:algo_RR}
  \end{minipage}
  &
  \begin{minipage}[t]{0.38\linewidth}
    \begin{algorithmic}[1]
        \Require{Trial vectors $\Psi, \ttH\Psi,\ttS\Psi$.}
        \Ensure{$\Lambda,\Psi,\ttH\Psi,\ttS\Psi$.}
        \State $H_{\Psi} =\Psi^\top (\ttH\Psi)$ \hfill \gemm
        \State $S_{\Psi} = \Psi^\top (\ttS\Psi)$ \hfill \gemm
        \State $H_{\Psi}X = \Lambda S_{\Psi}X$ \hfill \hegvd
        \State $\Psi\gets \Psi X$ \hfill \gemm
        \State $\ttH\Psi\gets (\ttH\Psi) X$ \hfill \gemm
        \State $\ttS\Psi\gets (\ttS\Psi) X$ \hfill \gemm
    \end{algorithmic}
    \InnerAlgorithmCaption{Matrix-free Rayleigh-Ritz procedure.}{fig:algo_dft}
  \end{minipage}
   \\ \hline
  \end{tabular}
\end{algorithm}

\paragraph{Matrix-free application of the Hamiltonian}

Unlike sparse real-space Hamiltonians, the plane-wave Hamiltonian is formally dense. To avoid assembling a dense matrix, \abinit implements a \textit{matrix-free} application of the plane-wave Hamiltonian. Matrix-free algorithms save memory by applying the Hamiltonian to a wave function (or any input column vector) without explicit allocation of any memory for storing the Hamiltonian matrix. Matrix-free algorithms are well-suited to plane-wave DFT because the number of wanted eigenpairs is much smaller than the number of plane-wave basis elements (typically few percents). For instance, for medium to large systems the order of magnitude is typically 10k-30k sought eigenpairs for 100k-1M plane waves. Moreover, as the plane-wave Hamiltonian is dense, explicit storage would scale quadratically with the number of plane waves. Matrix-free methods reduce memory requirements to linear scaling in the number of plane waves. In plane-wave basis, applying the Hamiltonian operator to an arbitrary coefficient vector involves FFTs and local multiplications. The matrix-free version of the Rayleigh-Ritz procedure implemented in \abinit is illustrated in Algorithm~\ref{fig:algo_dft}, where $\ttH\Psi$ and $\ttS\Psi$ denote the matrices obtained after matrix-free application of operators $H$ and $S$ to $\Psi$.

\paragraph{Subspace iteration}

\abinit implements two main classes of algorithms for constructing the trial subspace, using two different approaches. The first one is \textit{vector-oriented} and partitions the wave function into blocks of vectors to apply a block resolution, such as the Locally Optimal Block Preconditioned Conjugate Gradient (\lobpcg) \cite{knyazev2001,BOTTIN2008329}, while the second one is \textit{spectrum-oriented} and filters the eigenvalue spectrum using analytical properties of polynomials to amplify wanted spectral intervals, such as \chebyshevfiltering \cite{zhou2006,LEVITT201598} (using Chebyshev polynomials). All steps of these two families of algorithms are offloaded to GPUs allowing to keep the wave function in GPU-resident memory. Algorithms~\ref{algo:batched_lobpcg} and \ref{algo:batched_filter} illustrate the block structure used in the two GPU-ported eigensolvers when applying elementary operations on the wave function.

\begin{algorithm}[H]
\centering
\footnotesize
\setlength{\arrayrulewidth}{0.3pt} 
\setlength{\tabcolsep}{4pt}        

\begin{tabular}{|c|}
\hline

\begin{minipage}[t]{0.9\linewidth}
    \begin{algorithmic}[1]
        \Require{$k$ line minimizations, $M$ starting linearly independent electronic bands in $\Psi$ partitioned into $b$ blocks.}
        \Ensure{$\Lambda,\Psi$ solution to $H\Psi = \Lambda S\Psi$.}
        \For{$\Psi_i \in \Psi_1,\ldots,\Psi_b$}
        \State $\Psi_i \gets \text{RefineSubspaceBlock}(\Psi_1,\ldots,\Psi_i)$\hfill\textit{(deflation)}
        \State $\Lambda_i,\Psi_i\gets\text{ExtractEigenvectorsBlock}(\Psi_i)$\hfill\textit{(H app.,block~solver=RR)}
        \State Initialize conjugate vectors to $\Psi_i$ and $W_i$ as $P_i = 0$.
        \For{$n=1,\ldots,k$}
        \State $W_i \gets \text{ComputeResidual}(\Lambda_i,\Psi_i)$
        \State $W_i \gets \text{RefineSubspaceBlock}(\Psi_1,\ldots,\Psi_{i-1},W_i)$\hfill\textit{(deflation)}
        \State $\Lambda_i,\Psi_i,P_i\gets\text{ExtractEigenvectorsBlock}(\Psi_i,W_i,P_i)$\hfill\textit{(H app.,block~solver=RR)}
        \EndFor
        \EndFor
        \State $\Lambda,\Psi\gets\text{ExtractEigenvectors}(\Psi)$\hfill\textit{(global solver=RR)}
    \end{algorithmic}
    \footnotesize
    \InnerAlgorithmCaption{\lobpcg with $b$ blocks and $k$ line minimizations.}{algo:batched_lobpcg}
\end{minipage}

\\ \hline
\begin{minipage}[t]{0.9\linewidth}
    \begin{algorithmic}[1]
        \Require{$k$ polynomial degree, $M$ starting linearly independent electronic bands in $\Psi$.}
        \Ensure{$\Lambda,\Psi$ solution to $H\Psi = \Lambda S\Psi$.}
        \State Rayleigh quotients initialization.\hfill\textit{(H app.)}
        \For{$n=1,\ldots,k$}
        \State $\Psi \gets \text{SpectralFiltering}(\Psi)$\hfill\textit{(H app.)}
        \EndFor
        \State $\Lambda,\Psi\gets\text{ExtractEigenvectors}(\Psi)$\hfill\textit{(global solver=RR)}
    \end{algorithmic}
    \InnerAlgorithmCaption{Spectral polynomial filtering of degree $k$.}{algo:batched_filter}
\end{minipage}
\\ \hline
\end{tabular}
\end{algorithm}

Comparing Algorithms~\ref{algo:batched_lobpcg} and \ref{algo:batched_filter}, we observe that Hamiltonian application (H app.) and Rayleigh-Ritz (RR) procedures are common to both types of algorithms, but applied to different subspaces. We also notice that polynomial filtering eliminates inter-block dependencies, contrary to vector-oriented \lobpcg that includes inter-block dependence. Finally, a global solver call based on the Rayleigh-Ritz procedure is common in both methods. 

The LOBPCG implementation is identical to the first version in \abinit \cite{BOTTIN2008329}, which is similar to the original implementation \cite{knyazev2001,PETSc}. In our case the set of eigenpairs is divided into several blocks, which are treated sequentially.
The current block depends on the previous ones because of \textit{deflation}, which computes orthogonal vectors with respect to previous blocks.
Another difference is the Rayleigh quotients initialization is done by a complete Rayleigh-Ritz step. If a single block is used, the global Rayleigh-Ritz step at the end of \lobpcg is not needed, as in the original implementation.

\subsection{Parallel implementation design}\label{sec:paral_design}

We focus on the parallel implementation of iterative solvers based on the main computational kernels. These operations and their respective notation are (where $n$ and $m$ denote arbitrary integers):
\begin{itemize}
    \item $S$-deflation that corresponds to computing the orthogonal component of $m$ bands expressed in $n$ plane-wave coefficients with respect to the vector space generated by $m'$ orthogonal bands under the $S$-inner product, denoted by $\sdeflate(n,m,m')$,
    \item $S$-orthogonalization of $m$ bands expressed in $n$ plane-wave coefficients, performed by the matrix-free routine $\sortho(n,m)$, computed in practice using Cholesky factorization,
    \item Rayleigh-Ritz procedure to extract $m$ eigenvectors expressed in $n$ plane-wave coefficients, performed by the matrix-free routine $\rr(n,m)$ given by Algorithm~\ref{fig:algo_dft},
    \item Application of the Hamiltonian operator $H$ (and of the PAW overlap matrix $S$) to $m$ vectors expressed in $n$ plane-wave coefficients, performed by the matrix-free routine $\texttt{HX}(n,m)$ (and by $\texttt{HX\_SX}(n,m)$),
    \item Application of the inverse overlap operator $S$ to $m$ vectors expressed in $n$ plane-wave coefficients, using a Woodburry formulation, as explained in~\cite{LEVITT201598}, performed by the matrix-free routine $\texttt{Sinv}(n,m)$.
\end{itemize}

From now on, if $p$ is the number of parallel processes, let us define $m_p=M/p$ and $n_p=N/p$. For a matrix $A$ with $N$ rows and $M$ columns, we denote $A_i^p$ the submatrix containing all columns of $A$ but only the $n_p=N/p$ rows with indices $1+(i-1)n_p,\dots,in_p$. Noting $B$ a matrix with $N$ rows and $M'$ columns, we have $A^\top B=\sum_p (A_i^p)^\top B_i^p$, which shows that the product $A^\top B$ can be parallelized over $p$ processess doing matrix products with a cost of $\propto n_p MM'$ operations on each process and the final result is recovered by a reduction (\mpir) of $MM'$ matrix elements from all processes. This parallelization scheme distributes the work depending on rows size, and communicates data depending on column sizes only. It is then particularly efficient if $N\gg M$ and $N\gg M'$. We used this strategy to parallelize \sdeflate, \sortho and \rr matrix-free algorithms with parallel distribution of rows as follows:

\begin{algorithm}[H]
  \centering
  \footnotesize
  \setlength{\arrayrulewidth}{0.3pt} 
  \setlength{\tabcolsep}{4pt}        
  \begin{tabular}{|c|c|c|}
  \hline
  \begin{minipage}[t]{0.28\linewidth}
    \begin{algorithmic}[1]
        \Require{$\Psi,\Psi_0,\ttS\Psi_0$.}
        \Ensure{$\Psi$.}
        \State $S_{\Psi_0,\Psi} = (\ttS\Psi_0)^\top \Psi$ 
        \Statex \hfill \gemm+\mpir
        \State $\Psi\gets \Psi - \Psi_0 S_{\Psi_0,\Psi}$ \hfill \gemm
        \vspace{8ex}
    \end{algorithmic}
    \footnotesize
    \InnerAlgorithmCaption{\sdeflate}{fig:algo_Sdeflate}
  \end{minipage}
  &
  \begin{minipage}[t]{0.32\linewidth}
    \begin{algorithmic}[1]
        \Require{Trial vectors $\Psi,\ttH\Psi,\ttS\Psi$.}
        \Ensure{$\Psi,\ttH\Psi,\ttS\Psi$.}
        \State $S_{\Psi} = \Psi^\top (\ttS\Psi)$ \hfill \gemm +\mpir
        \State $S_{\Psi} = U^\top U$ \hfill \potrf
        \State $\Psi\gets \Psi U^{-1}$ \hfill \trsm
        \State $\ttH\Psi\gets (\ttH\Psi) U^{-1}$ \hfill \trsm
        \State $\ttS\Psi\gets (\ttS\Psi) U^{-1}$ \hfill \trsm
        \vspace{3ex}
    \end{algorithmic}
    \footnotesize
    \InnerAlgorithmCaption{\sortho}{fig:algo_Sortho}
  \end{minipage}
  &
  \begin{minipage}[t]{0.33\linewidth}
    \begin{algorithmic}[1]
        \Require{$\Psi, \ttH\Psi,\ttS\Psi$.}
        \Ensure{$\Lambda,\Psi,\ttH\Psi,\ttS\Psi$.}
        \State $H_{\Psi} =\Psi^\top (\ttH\Psi)$ \hfill \gemm +\mpir
        \State $S_{\Psi} = \Psi^\top (\ttS\Psi)$ \hfill \gemm +\mpir
        \State $H_{\Psi}X = \Lambda S_{\Psi}X$ \hfill \hegvd
        \State $\Psi\gets \Psi X$ \hfill \gemm
        \State $\ttH\Psi\gets (\ttH\Psi) X$ \hfill \gemm
        \State $\ttS\Psi\gets (\ttS\Psi) X$ \hfill \gemm
    \end{algorithmic}
    \InnerAlgorithmCaption{\rr}{fig:algo_RRparal}
  \end{minipage}
   \\ \hline
  \end{tabular}
\end{algorithm}

The only communications are MPI-reductions (\mpir) done in the first matrix-matrix multiplications, as in the replicated approach of \cite{LUPOPASINI2020102703}. Therefore, all calls to LAPACK routines are done without MPI parallelism. In some calls of \rr in \lobpcg, there is an additional step which is a linear combination of output vectors to compute conjugate vectors of $\Psi$ \cite{BOTTIN2008329,knyazev2001,PETSc}. This is identical to the parallel scheme used in \cite{LUPOPASINI2020102703}. Note that we also evaluated distributed multi-GPU eigensolvers based on cuSolverMP \cite{cusolvermp} and ELPA \cite{marek2014elpa}. For the relatively small projected matrices considered here, however, communication and distribution overheads outweighed the benefits of parallelism, making replicated solves more efficient. We reached the same conclusion for CPUs when comparing \lapack and ScaLAPACK. Thus, replicated solves are not specific to GPUs, as \abinit also employs this strategy in its MPI+\openmp CPU implementation.

Small reduction operations (\mpisr) are also needed to compute the norms of row-wise distributed residuals. The $M$ dot products (or $M/b$ in the loop over blocks) are obtained with $\propto n_pM$ operations ($n_pM/b$) and a reduction over all processes. In \chebyshevfiltering, there is also a small reduction to gather the minimum and maximum of Rayleigh quotients $\lambda_i=(H\psi_i)^\top\psi_i/((S\psi_i)^\top\psi_i)$ for ($i=1,\ldots,M/p$) where wave functions are column-wise distributed.

Algorithms~\ref{alg:lobpcg_k} and \ref{alg:chebfi_k} in \ref{sec:parallel_algos_detail} show the parallel diagonalization algorithms, in which every steps using MPI communications are explicitly indicated. The Chebyshev polynomial spectral filtering algorithm employs an auxiliary kernel to efficiently compute the Hamiltonian polynomial applied to wave functions via a recurrence relation. It involves operations of the form $X\gets aX + Y$, for $X$ and $Y$ of size $n\times m$, performed by the routine $\texttt{axpy}(n,m)$ of level-1 BLAS. Parallel processes communicate only at the end of the subspace iteration of length $k$. Note that the loop over the number of blocks as well as the loop over the number of inner iterations are intrinsically sequential.

The main difference between the algorithms is that \lobpcg{} makes intensive use of the Rayleigh-Ritz (RR) procedure (and orthogonalizations) by calling it within each subspace iteration plus a global one ($b$ calls of $(M/b)$-sized RR, $b$ calls of $2(M/b)$-sized RR, $b(k-1)$ calls of $3(M/b)$-sized RR and one call of $M$-sized RR), whereas \chebyshevfiltering{} makes only a global one ($M$-sized RR). As we demonstrate in the following, the Rayleigh-Ritz and orthogonalization procedures scale poorly with respect to the number of bands and do not fully benefit from efficient parallelization.

\subsection{Floating-point operation count}\label{sec:flop_count}

To quantify the computational cost of each algorithm, we estimate the floating-point operation (FLOP) count using the following theoretical models. We assume that $k$ is the maximal number of inner iterations (gradient descent or polynomial filtering degree) and $b$ is the number of blocks in \lobpcg algorithm. Let $N_{\text{projs}}$ be the number of projectors involved in the overlap operator. The steps of diagonalization algorithms, as appearing in Algorithms~\ref{alg:lobpcg_k} and \ref{alg:chebfi_k}, admit the following theoretical cost estimates. Let $m$ be an arbitrary number of bands. The routine $\rr(n_p,m)$ approximately has a cost of $O(m^3+n_pm^2)$ FLOP (\gemm and \hegvd dense \lapack solver for the problem on a subspace of dimension $m$) \cite{doi:10.1137/070688778}. The routine $\sortho(n_p,m)$ is also dominated by cubic scaling with respect to $m$, as its implementation involves the multiplication $X^\top (\ttS X)$ (\gemm routine), the Cholesky factorization of a Hermitian positive-definite matrix (\texttt{potrf} routine) and the triangular matrix solver (\trsm routine). The routine $\sdeflate(n_p,m,m')$ has a cost of $O(n_pmm')$ FLOP as it involves \gemm routines only. The cost of $\texttt{HX}(N,1)$ is $O(N\log N+N_{\text{projs}}N)$ FLOP due to FFT \cite{LEVITT201598,d3ea2d52-5ab2-3128-8b80-efb85267295d}. Then $\texttt{HX}(N,m)$ has $m$ times the cost of $\texttt{HX}(N,1)$ due to batching over bands. Finally, according to \cite{LEVITT201598}, the iterative algorithm used to apply $S^{-1}$ has a cost of $O(N_{\text{projs}}^2)\ll O(N_{\text{projs}}N)$, so the total cost of $\texttt{Sinv}(N,1)$ applied on one band is $O(N_{\text{projs}}N)$.

Now, using parallelization, the cubic scaling of Rayleigh-Ritz and of orthogonalization is not affected when
parallelizing over plane waves. On the other hand, by parallelizing the Hamiltonian and the $S,S^{-1}$ applications over
bands, we ideally gain a factor $1/p$ when using $p$ parallel processes. Additionally note that multithreading within a single MPI task performs poorly for the eigensolver (\texttt{hegvd}) and the Cholesky factorization (\texttt{potrf}).

Overall, the Rayleigh-Ritz procedure and $S$-orthogonalization exhibit cubic scaling with the number of bands, while the Hamiltonian application and $S$-inversion scale linearly with the number of bands. This theoretical estimate reveals that the final Rayleigh-Ritz step scales poorly in both methods and is expected to represent a computational bottleneck for systems with a large number of bands. Thus, the number of bands emerges as a limiting dimension for these operations.

\subsection{\mpi collective communication operations}\label{sec:mpi_count}

This discussion is independent of the presence of a GPU and focuses only on comparing \lobpcg and \chebyshevfiltering,
all else being equal. We will examine the data transfers described in Algorithms~\ref{alg:lobpcg_k} and
\ref{alg:chebfi_k} in order to obtain a comparative estimate of the communication call counts and buffer sizes. The \mpi communication cost, in terms of latency and bandwidth usage, of a single point-to-point message containing $\mathsf n$ bytes can be estimated using the cost model ``$\alpha+\mathsf n\beta$''~\cite{doi:10.1177/1094342005051521}, where $\alpha$ is the latency (s) (independent of the message size) and $\beta$ is the inverse bandwidth. The cost of a MPI collective is obtained by composing this model according to the collective communication algorithm and the number of participating processes. Since the particular implementation may depend on the MPI library, message size, and network topology, we report the number of collective calls and the corresponding memory-buffer size per MPI process.

The most expensive communication operation is the \alltoall (\mpia), introduced in Section~\ref{sec:MPI_between_GPU}, which switches the data distribution from row-wise to column-wise or vice versa. It involves $p$ processes each one sending $n_p$ rows and $M/b$ columns to $p-1$ other processes, while the inverse operation sends $N$ rows and $m_p/b$ columns. With double-precision complex buffers, the message size sent by one process is $16NM/(pb)$ bytes in \lobpcg and $16NM/p$ bytes in \chebyshevfiltering. In \lobpcg, back and forth MPI transpositions repeat $k+1$ times per block, once in each of the $k$ CG iterations, whereas in \chebyshevfiltering they occur only once. \lobpcg invokes $b(k+1)$ times as many \alltoall collectives as \chebyshevfiltering. Table~\ref{tab:mpi_comms_alltoall} summarizes the \alltoall counts and buffer sizes.

\begin{table}[H]
    \footnotesize
    \centering
    \begin{tabular}{l|r|r}\toprule
      all-to-all (\mpia) & \texttt{\lobpcg}$(N,M,b,k)$   &  \texttt{ChebFi}$(N,M,k)$ \\ \midrule
          Number of collectives &   $4 b(k+1) $ &  $4$  \\
          Buffer size (bytes) &   $16 NM/(pb)$ & $16 NM/p$  \\ \bottomrule
    \end{tabular}
    \caption{Number \alltoall collectives and buffer size per collective per MPI process for the \lobpcg and \chebyshevfiltering (abbreviated as \texttt{ChebFi}) parallel algorithms. Note there are 4 transpositions per Hamiltonian application: $\Psi$ (forward), $\Psi$ (backward), $\ttH\Psi$ (backward), $\ttS\Psi$ (backward).}
    \label{tab:mpi_comms_alltoall}
\end{table}

The other important communication operation is the reduction (\mpir) in the computation of matrix products $A^\top B$ with row-wise distributed data, as explained in Section~\ref{sec:paral_design}. In \lobpcg, this happens during the loop over blocks in calls to $\sdeflate, \sortho$ and $\rr$ with message sizes proportional to $16M^2/b^2$ sent by one process to $p-1$ other processes. It also occurs at the end of \lobpcg and \chebyshevfiltering algorithms in the global $\sortho$ and $\rr$, with message sizes equal to $16M^2$. This kind of MPI reduction costs are summarized in Table~\ref{tab:mpi_comms_reduction}. 

\begin{table}[H]
    \footnotesize
    \centering
    \begin{tabular}{l|r|r|r|r|r}\toprule
      reduction (\mpir)  & \multicolumn{3}{c|}{\texttt{\lobpcg}$(N,M,b,k)$ -- block part} & \multicolumn{2}{c}{global} \\ 
      \cmidrule(lr){2-4}
      \cmidrule(lr){5-6}
                         & \sdeflate        & \sortho & \rr & \sortho & \rr \\ \midrule
          Number of collectives    & $(b-1)(k+1)$   & $b(k+1)$  & $2b(k+1)$ & $1$ & $2$ \\
          Buffer size (bytes)      & $\propto 16 M^2/b^2$ & $\propto 16 M^2/b^2$ & $\propto 16 M^2/b^2$ & $16 M^2$ & $16 M^2$
    \\ \bottomrule
    \end{tabular}
    \caption{Number of reduction collectives and buffer size per collective per MPI process in the \lobpcg loop over blocks, and for the global \sortho and \rr at the end of both \lobpcg and \chebyshevfiltering algorithms. For \lobpcg, we note that the global reduction occurs only if $b>1$.}
   \label{tab:mpi_comms_reduction}
\end{table}

At last, \mpisr operation costs are small. Indeed in the computation of residual norms the message size is $16M$ (or $16M/b$ in the loop over the blocks) and a message of only $16$ bytes is needed to gather the maximum and minimum of Rayleigh quotients in \chebyshevfiltering.

The theoretical estimates show that number of communication calls in \lobpcg scale linearly with the total number of blocks; thus, we expect \lobpcg to suffer from communication overhead when dealing with large systems or a larger number of blocks. It is important to keep in mind that \mpi communications are performed between GPUs if GPU-aware MPI is supported.

\subsection{Arithmetic intensity}\label{sec:arithm_intensity}

The arithmetic intensity $I$ is defined as the ratio of computational work $W$ to total memory traffic (number of read/write requests $\times$ bytes per request) $Q$:
$$I = \frac{W}{Q}.$$
The work is measured in FLOP. Memory traffic represents the total number of bytes transferred during read/write requests
to DRAM, on a single \mpi process during GPU kernels execution. Here, we aim to estimate the arithmetic intensity for
kernels between two consecutive \alltoall operations.

Efficient algorithms are designed to maximize arithmetic intensity per \mpi process. The parallel \lobpcg and \chebyshevfiltering algorithms handle \mpi GPU-to-GPU communications differently for a fixed number $k$ of inner
iterations during the subspace iteration step. Indeed, \lobpcg applies the Hamiltonian $k+1$ times per block
(i.e., ``$(k+1)\,\times\,$Comm.--$H$ application--Comm.'' per block), resulting in a total of $4b(k+1)$ \alltoall operations. In contrast, \chebyshevfiltering applies only 4 global \alltoall
operations in total (i.e., ``Comm.--$H^{k+1}$ application--Comm.''). Intuitively, \chebyshevfiltering maximizes
arithmetic intensity by applying the Hamiltonian $k+1$ times more per \mpi data transposition than \lobpcg.

Indeed, an estimate of the arithmetic intensity of the GPU kernels involved in Hamiltonian applications, for a single data movement, is
\begin{align*}
    I(\text{\lobpcg-}\texttt{HX}) &= \frac{\text{FLOP}(\texttt{HX}(N,m_p/b))}{Q/b},\\
    I(\text{\chebyshevfiltering-}\texttt{HX}) &= \frac{(k+1)\times\text{FLOP}(\texttt{HX}(N,m_p))}{Q},
\end{align*}
where $Q$ is the byte count for storing a complex wave function in column-distribution of size $N\times m_p$ on the GPU
in double precision, yielding $Q=16Nm_p$ bytes of input and output memory buffers. Since the Hamiltonian application is parallel in the number of bands, $\text{FLOP}(\texttt{HX}(N,m_p/b))=\text{FLOP}(\texttt{HX}(N,m_p))/b$ and finally:
\begin{align*}
  I(\text{\chebyshevfiltering-}\texttt{HX}) = (k+1) I(\text{LOBPCG-}\texttt{HX}).
\end{align*}
This shows that \chebyshevfiltering maximizes the arithmetic intensity of the Hamiltonian application operation, which is well suited for GPUs thanks to its batched execution.

\section{Methods and programming models}\label{sec:methods}

This section presents specific strategies used for offloading computationally expensive parts of the code on GPUs. More details on how low-level GPU libraries communicate with the \abinit code are provided. Basically we use the following hierarchy of operations from a development view-point:
\begin{enumerate}[label=(\roman*)]
    \item \textit{Directive-based programming}. The\textit{} way data transfers between host and device are executed, the way we interface with GPU libraries at a programming level.
    \item \textit{Low-level operations}. Interfaces strictly connected to vendor libraries. Wrappers to libraries. Mathematical operations (matrix-matrix multiplications, FFTs, eigensolvers) are mapped to GPU libraries.
    \item \textit{Abstraction layer}. Implements data-structures adapted to the commonly used objects in \abinit, namely the wave-function. The architecture is defined using a keyword and for every architecture there is a wrapper to its implementation. Boolean options are used to control if an operation is executed on GPU or CPU.
\end{enumerate}

\subsection{\openmpoffload}

We employ the \openmpoffload programming model for GPU acceleration, using directives introduced in \openmp API version 5.0~\cite{OpenMP5.0}. \openmp handles device workload distribution, memory management and host-device data transfers via \texttt{target} directives,  with execution controlled by the
\texttt{OMP\_TARGET\_OFFLOAD} environment variable.

Computations are performed on the device within an \openmptarget region. We selected this programming model for \begin{enumerate*}[label=(\roman*)]
\item its directive-based simplicity, which facilitates porting with minimal code rewriting; \item its excellent portability across platforms; \item its limited compilation dependencies; and \item its simplified memory management, which allows allocations and frees to be handled directly in Fortran. \end{enumerate*} Together, these aspects enhance the code's readability and maintainability over time. We systematically apply the \texttt{COLLAPSE} directive for batch loop processing, enabling straightforward GPU persistent memory management. The model is also vendor-agnostic; we validated \openmptarget with \cuda and \nvhpc compiler on \nvidia GPUs (A100, H100, H200) and with \rocm and \cray compiler on \amd GPUs (MI250, MI300). We validated unified memory usage on NVIDIA GH100 and GH200, which is greatly simplified in the \openmptarget paradigm.

\subsection{Low-level operations}\label{sec:lowlevelops}

This section summarizes the GPU implementation of elementary low-level operations. Our current approach primarily employs vendor libraries for maximum portability, avoiding custom GPU kernels. We rely on backend libraries from \cuda Toolkit \cite{cuda_manual} for \nvidia GPUs, and \rocm \cite{rocm_manual} for \amd GPUs. Calls to these vendor library functions are encapsulated in wrappers, making them transparent to the overlying software layer.

Firstly, batching is employed to compute the FFT of multiple wave functions stored in an array through a single call,
instead of applying the FFT iteratively to individual vectors. The plane-wave cut-off energy $E_{\text{cut}}$ determines
the grid size, and consequently the size of the FFT, while the number of bands treated per MPI task determines the
number of FFTs, consequently the batch size. This process is used when computing the local part of the Hamiltonian
operator. Since this operation involves all plane waves, the data must first be rearranged into the specific layout
described in Section~\ref{sec:MPI_between_GPU}. The vendor libraries \cufft \cite{cufft} and \rocfft \cite{rocfft} are then used in a batched and optimized mode to maximize reuse of precomputed data whenever possible.

Secondly, we specifically address dense matrix-matrix multiplications (i.e. \gemm operations) and associated routines critical for GPU-accelerated Hamiltonian. Fortran \blas \gemm calls are replaced with equivalents from \cublas \cite{cublas} and \rocblas \cite{rocblas}. Notably, the matrix-free plane-wave Hamiltonian application leverages level-3 \blas operations for the calculation of the nonlocal contribution, as applications to different bands are independent—favoring matrix-matrix over matrix-vector multiplications.

Thirdly, for solvers, we employ \cusolver \cite{cusolver} and \rocsolver \cite{rocsolver} to tackle the generalized eigenvalue problem within the Rayleigh-Ritz procedure using dense eigensolvers for Hermitian matrices. This computational step proves critical, as the Rayleigh-Ritz procedure represents a major bottleneck (see Section~\ref{sec:flop_count}). Consequently, the performance of the \lapack routine \hegvd—which varies significantly across implementations—plays a pivotal role.

Finally, an explicit code refactoring enables batched kernel execution on GPUs for applying the non-local operator, expressed as $V_{\rm nonloc}=\sum_{R,i,j}{\ket{p^R_i}D^R_{ij}\bra{p^R_j}}$, where the $\ket{p^R_i}$ are projectors depending on atom $R$ and local basis index $1\leq i\leq n^R_{lmn}$. Ultimately,  application requires matrix-matrix multiplications involving $P=\braket{g|p^R_i}$ (projectors in the plane-wave basis), $C_{\rm{proj}}=\braket{p^R_i|\Psi}$ (projected wave functions), and $D^R_{ij}$. Since the local basis size $n^R_{lmn}$ varies by atom type, batching these matrix-matrix multiplications requires subtleties. This was addressed by refactoring the code: loops were reordered to process atomic types sequentially first, while grouping wave functions into batches. The \texttt{gemmBatchedStrided} kernel, available in \cublas and \rocblas, efficiently computes groups of matrix-matrix products on the GPU. Refactoring additionally involved transforming some function calls that previously passed Fortran array slices. Avoiding slices improves control over memory layout and exposes greater data parallelism.

\subsection{Abstraction layer}\label{sec:abstraction}

\begin{figure}[H]
    \centering
    \includegraphics[width=0.55\linewidth]{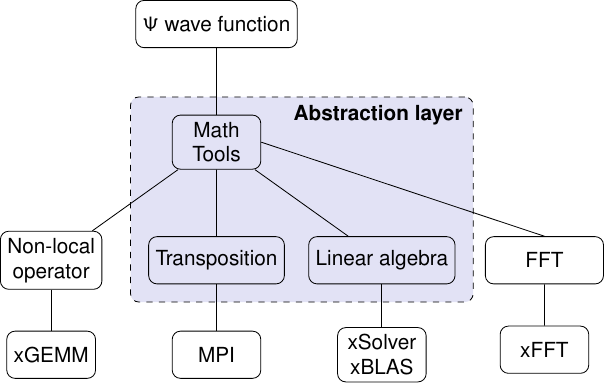}
    \caption{Hierarchy from the high-level wave function to the low-level GPU libraries (x=``cu'', ``roc'') linked through the \abinit abstraction layer. FFT and non-local operator will be included in the abstraction layer in a future work.}
    \label{fig:xgtools}
\end{figure}

An abstraction layer enables algorithms to interface with low-level operations from a software development perspective, as illustrated in Figure~\ref{fig:xgtools}.

This abstraction layer provides Fortran types offering a unified, high-level interface for handling distributed matrices and vectors, performing algebraic operations, and solving generalized eigenvalue problems. These types operate on blocks of row (or column) vectors forming contiguous memory batches, with a data structure compatible with GPU architectures and support for CPU/GPU memory pointers. Designed for clarity and simplicity—especially for \abinit developers—the interface hides the underlying complexity and diversity of algebraic structures used in wave-function manipulations across linear algebra libraries.
The abstraction layer facilitates easy extension by users through additional functionalities. The \textit{Tools} framework supports matrix-free Hamiltonian applications by manipulating objects $X$ (the wave function) and $AX$ (its application under operator $A$), without needing explicit knowledge of the matrix $A$. The module manages all memory allocations/deallocations of workspaces  needed in diagonalization algorithms.

The abstraction layer also provides an interface between wave functions and \mpi routines. It enables efficient, transparent conversions between different wave function representations and their \mpi distributions (see Section~\ref{sec:MPI_between_GPU}), while managing contiguous memory layouts. All required communications are handled transparently—hidden from developers via simple routine options—supporting both row and column distributions. Sub-distributions via sub-communicators are straightforward to implement, facilitating additional parallelism across matrix blocks when needed.

The abstraction layer enables efficient execution of linear algebra operations while seamlessly interfacing with \blas and \lapack implementations for dense complex-valued matrices.

\section{Performance}\label{sec:perfs}

This section presents performance benchmarks and numerical results for the \abinit GPU port. We focus on achieved speedup, MPI communications between GPUs, energy efficiency and roofline models across \nvidia and \amd GPUs. All CPU versus CPU-GPU comparisons are made on a per-node basis. We also examine parameter tuning of \lobpcg and \chebyshevfiltering algorithms for achieving good performance on GPU. Computer specifications are detailed in \ref{sec:gpu_specifs}. Machines from Table~\ref{tab:gpu_specifs} are abbreviated as \adastra (CINES) and \topaze (CCRT).

\subsection{Benchmark test cases}\label{sec:test_cases}

\textit{Converged-SCF} benchmarks (Section~\ref{sec:multiple_gpu}) are obtained from converged calculations for a 255-atom solid-liquid titanium interface, using $M=8192$ electronic bands, a plane-wave cutoff energy of 8 Ha corresponding to $N=34576$ plane waves, a single \kpoint and PAW pseudopotentials. The number of self-consistent iterations was fixed to 20 and the diagonalization algorithm under test is \chebyshevfiltering with polynomial degree equal to $k=6$. We use \abinit 10.8. These tests ran on \topaze (CCRT) with \nvidia GPUs (\cuda 12.4, \nvhpc 24.5, MKL 24.2) and on \adastra (CINES) with AMD GPUs (MI250X partition: \cray CPE 25.09, \rocm 7.2.0; GENOA partition: \cray CPE 24.07). For this test, the \abinit input parameters (\openmp threads, \blas tuning, etc.) were selected to achieve the best observed performance.

\textit{Single-step} benchmarks (Sections~\ref{sec:per_gpu_diago} and~\ref{sec:per_gpu_si}) are obtained from constant-Hamiltonian calculations for 320 atoms of \ce{Ga2O3}, using $M=1536$ electronic bands, a cutoff energy of 20 Ha corresponding to $N=6316$ plane waves, a single \kpoint and PAW pseudopotentials. We perform a single self-consistent field iteration and \lobpcg has a fixed number of $b=4$ blocks. These tests ran on \topaze (CCRT) using \abinit version 10.4.5 compiled with \cuda 12.9 and \nvhpc 25.7 on \nvidia GPUs. The GPU profiles where collected using \nvidia Nsight Compute 2025.2.1~\cite{nvidia_nsight_compute}. The hardware performance counters on \amd were collected with \abinit 10.8 using rocprofv3 from ROCprofiler-SDK 0.6.0 \cite{amd_rocprofiler_sdk_2025}, distributed with ROCm 6.4.3. In all rooflines \lobpcg has $b=4$ blocks and $k=4$ line minimizations; \chebyshevfiltering has $k=10$ polynomial degree.

\subsection{Converged-SCF benchmarks: CPU versus CPU-GPU nodes}
\label{sec:multiple_gpu}

Using \textit{converged-SCF} test case, we compare GPU-accelerated performance and energy consumption between \texttt{n} CPU-only nodes and \texttt{n} CPU-GPU nodes---a true ``node-to-node'' comparison---for \texttt{n} ranging from 1 to 8. Each CPU-only node is directly benchmarked against one hybrid CPU-GPU node in terms of execution time, GPU communication and energy consumption.

\begin{figure}[H]
    \centering
    \includegraphics[width=\textwidth]{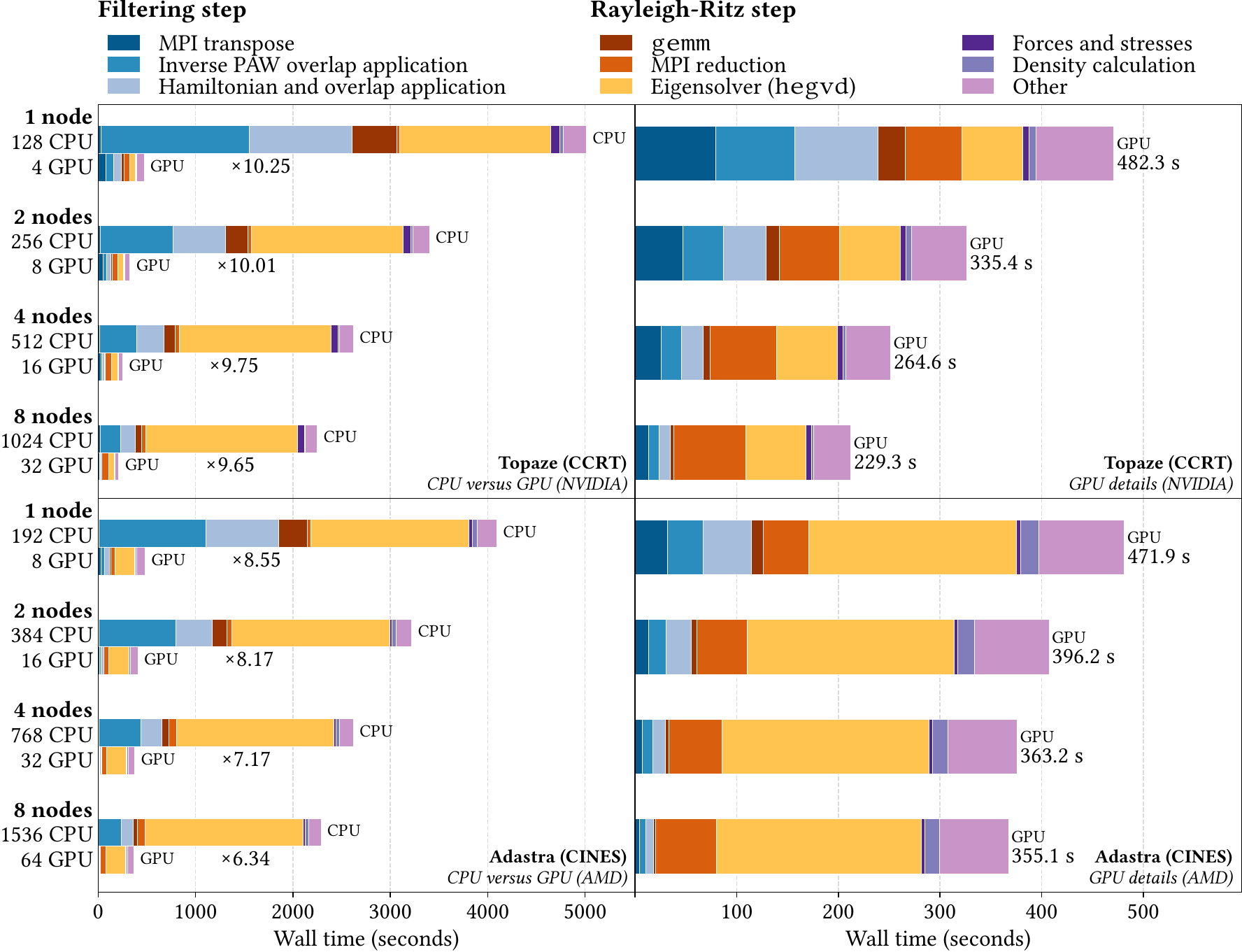}
    \caption{Execution times and GPU speedup for \chebyshevfiltering on a 255-atom Ti system (8192 electronic bands) over 20 SCF iterations. Results compare the performance between \topaze (TGCC, \nvidia GPUs) with \adastra (CINES, \amd GPUs), using the compiler and library versions specified in Section~\ref{sec:test_cases}. ``Filtering step'' includes \mpi transpositions (\mpia), matrix-free Hamiltonian, PAW overlap and inverse overlap applications (steps 1-9 Algorithm~\ref{alg:chebfi_k}); ``Rayleigh-Ritz step'' covers subspace diagonalization, further broken down into matrix-matrix products (\gemm), MPI reductions (\mpir) and subspace diagonalization (\hegvd) (Algorithm~\ref{fig:algo_RRparal}); remaining calculations include force and stress calculations, electronic density calculation and initialization-related operations (input file reading, I/Os, pseudopotential reading, and initial potential setup).}
    \label{fig:prel}
\end{figure}

Figure~\ref{fig:prel} illustrates GPU speedup for \chebyshevfiltering. \nvidia GPUs consistently deliver higher speedup factors than \amd GPUs. The Rayleigh-Ritz step dominates execution time and exhibits lower acceleration than the filtering step, which becomes negligible with GPU acceleration. The Rayleigh-Ritz step proves particularly poorly suited to \amd GPUs. Closer inspection of per-step timings reveals that the Rayleigh-Ritz portion increases with node count. Overall, filtering benefits substantially from acceleration, while Rayleigh-Ritz improves more modestly, particularly on \amd architecture. As shown in Figure~\ref{fig:prel}, this is related to the performance of the \hegvd routine called during the Rayleigh-Ritz step. We respectively used \cusolver 11.6.1.9 for \nvhpc 24.5 on \nvidia GPUs and \rocsolver 3.32.0 for \rocm 7.2.0 on \amd GPUs. Despite various improvements to the \rocm implementation of \hegvd \cite{rocsolver_rocm642}, the eigensolver in \nvidia largely outperforms the one on \amd. Globally the speedup is excellent: 2 GPU nodes outperform 8 CPU nodes, enabling substantial resource savings.

Figure~\ref{fig:prel} also reveals insights into strong scaling. GPU scaling efficiency lags behind CPU performance as node count grows, consistent with \textit{Amdahl’s law}. In practice, thanks to this limited scalability, fewer GPUs are needed compared to CPU-only scaling. Notably, the Rayleigh-Ritz \texttt{hegvd} component shows no acceleration with additional GPU nodes on both \topaze and \adastra, with its scaling factor remaining close to unity. On \topaze, the inverse PAW-overlap operation, the Hamiltonian and overlap applications, and the Rayleigh-Ritz \texttt{gemm} exhibit nearly ideal strong scaling. These operations scale less efficiently on \adastra, where the MPI transpose achieves the closest-to-ideal scaling. Moreover, while MPI transpose continues to benefit from increasing the number of nodes, MPI reduction exhibits poor strong scaling and accounts for an increasing fraction of the GPU execution time. This effect is particularly pronounced on \topaze, where MPI reduction represents nearly one-third of the total GPU runtime at eight nodes.

To examine GPU MPI communication, we performed calculations with a reduced number of electronic bands, $M=4096$, and compared them with the $M=8192$ calculations at the same node counts. At each node count, increasing $M$ from 4096 to 8192 increases the MPI-reduction time by approximately a factor of four on both \topaze and \adastra, consistent with the quadratic growth of the distributed matrices with the number of electronic bands (Table~\ref{tab:mpi_comms_reduction}). Over the same increase in $M$, the MPI-transpose time on \topaze approximately doubles at every node count, consistent with its nearly linear dependence on the number of bands (Table~\ref{tab:mpi_comms_alltoall}). On \adastra, however, the MPI-transpose time for $M=8192$ is approximately 6--10 times that measured for $M=4096$ at the corresponding node count. This indicates non-linear scaling of the MPI-transpose operation on \adastra, although the present measurements do not identify its underlying cause.

\begin{figure}[H]
    \centering
    \includegraphics[width=0.7\linewidth]{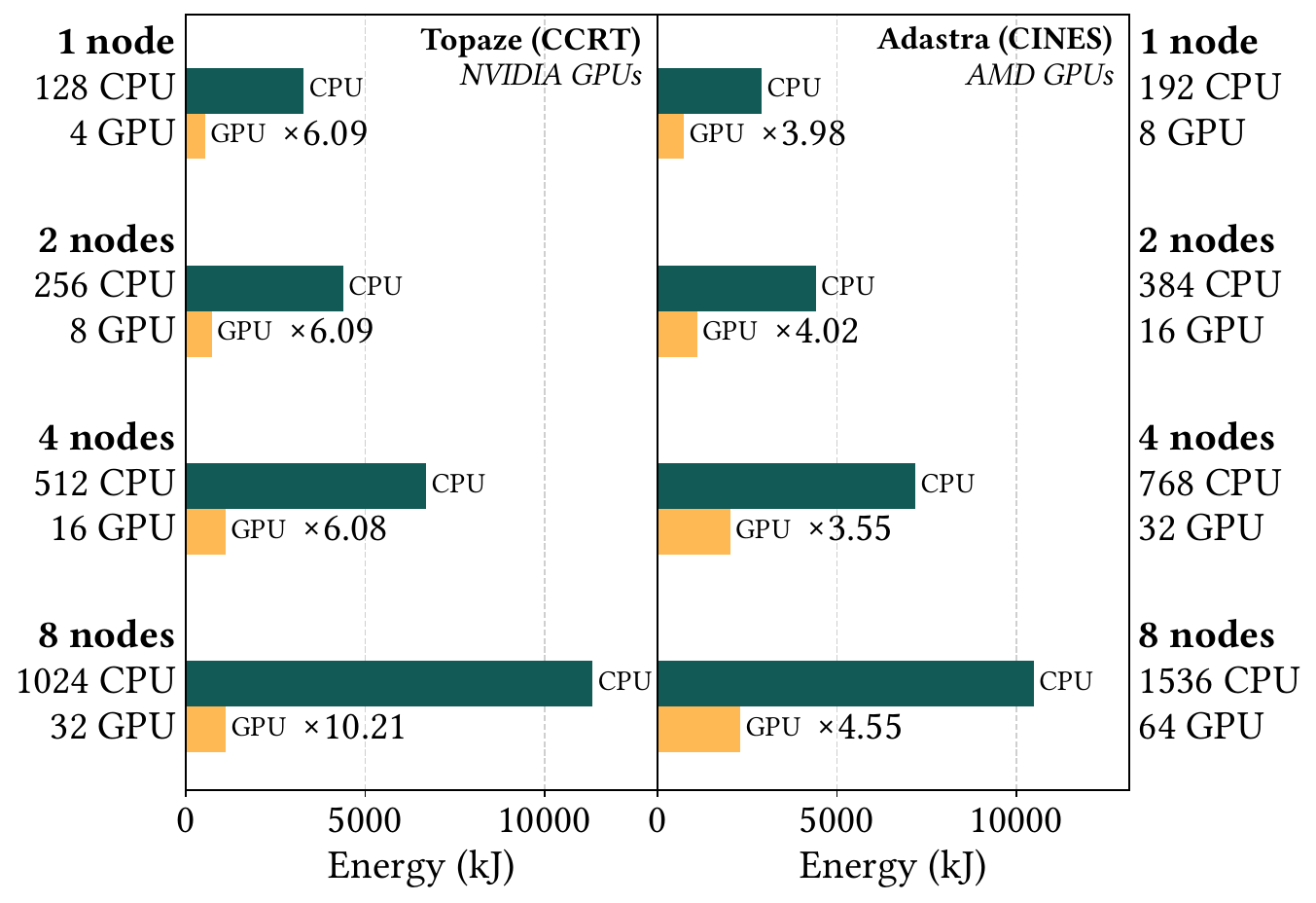}
    \caption{Energy consumption and savings for \chebyshevfiltering on a 255-atom Ti system (8192 electronic bands). Energy saving factors compare \topaze (CCRT, \nvidia GPUs) versus \adastra (CINES, \amd GPUs), with the same compiler and library versions specified in Section~\ref{sec:test_cases}. Energy measurements reflect compute-node power draw only (CPUs, accelerators, and DRAM), as reported by SLURM's energy accounting at job completion. Facility overhead (cooling, interconnect, storage) is not included; applying a typical Power Usage Effectiveness (PUE) of 1.05-1.10 for both systems (as reported in system documentation) would increase all values by at most 10\%. Energy measurements may differ between the two machines. Therefore, comparing only the energy saving factor is meaningful.}
    \label{fig:energy}
\end{figure}

To measure energy consumption, we used vendor-provided counters and tools, executed on the same hybrid nodes, by enabling/disabling GPU usage at the code level. Absolute energy values are therefore not comparable across vendors due to differing measurement methods. Figure~\ref{fig:energy} shows energy consumption on our multi-GPU test case. \nvidia GPUs on \topaze achieve higher saving factors than \amd GPUs on \adastra, which is directly related to the speedup factors observed on Figure~\ref{fig:prel}. Strong scaling reveals further differences. Energy consumption on \topaze remains nearly constant with additional GPU nodes, while on \adastra it increases and nearly doubles from 2 to 4 nodes. A detailed analysis reveals that the differences in energy gains are primarily due to the performance of the \lapack \hegvd routine used in the Rayleigh-Ritz step, where the \rocm library version 7.2.0 exhibits poor performance on AMD GPUs.

\subsection{Single-step benchmarks: GPU kernel rooflines}\label{sec:per_gpu_diago}

For a deeper analysis of GPU kernels performance, we gathered key metrics including compute performance and arithmetic intensity. These measures enable us to classify code sections as \textit{compute-bound} (already well-optimized) or \textit{memory-bound}, thereby revealing bottlenecks. These tests used a single-step benchmark: one SCF iteration with a fixed Hamiltonian on a 320-atom \ce{Ga2O3} cell ($M=1536$ electronic states).

The \textit{roofline model}~\cite{10.1145/1498765.1498785} is used to quantify the efficiency of a GPU kernel by measuring operations performed per byte of memory accessed. Here, we apply it to assess intra-GPU node overheads. We derive roofline limiting bounds from the GPU vendor’s peak specifications, here \nvidia A100 card~\cite{nvidia2020a100} and \amd MI250X \cite{CINES2026}. Measurements from actual executions reveal both compute-bound and memory-bound operations.

In practice, we collect measurements from hardware counters using NVTX/ROCTx markers to isolate code regions. The measured arithmetic intensity $I$ and performance $P$ is
\begin{equation*}
    I = \frac{\sum\,\texttt{FLOP}}{\sum\,\texttt{DRAM}}, \quad
    P=\frac{\sum\, \texttt{FLOP}}{\sum\, \texttt{time}}
\end{equation*}
where \texttt{FLOP} corresponds to FP64 operations. For \nvidia this is $\texttt{\cuda}+\texttt{Tensor}$ where \texttt{\cuda} measures total double-precision operations (dadd + dmul + 2$\times$dfma) on \cuda cores, \texttt{Tensor} measures total matrix-multiply-accumulate (mma) instructions on Tensor cores. \texttt{DRAM} measures kernel memory reads and writes in bytes (including uncoalesced DRAM access, cache effects and overhead from GPU runtime, unlike formula in Section~\ref{sec:arithm_intensity}). Finally \texttt{time} is the execution time. These quantities are deduced from metrics per GPU kernel execution, obtained with Nsight Compute and rocprofv3 reports, and then aggregated to obtain the total FLOP and bytes per code section. Profiling data were collected for MPI process 0 only.

\begin{figure}[H]
    \centering
    \includegraphics[width=0.98\linewidth]{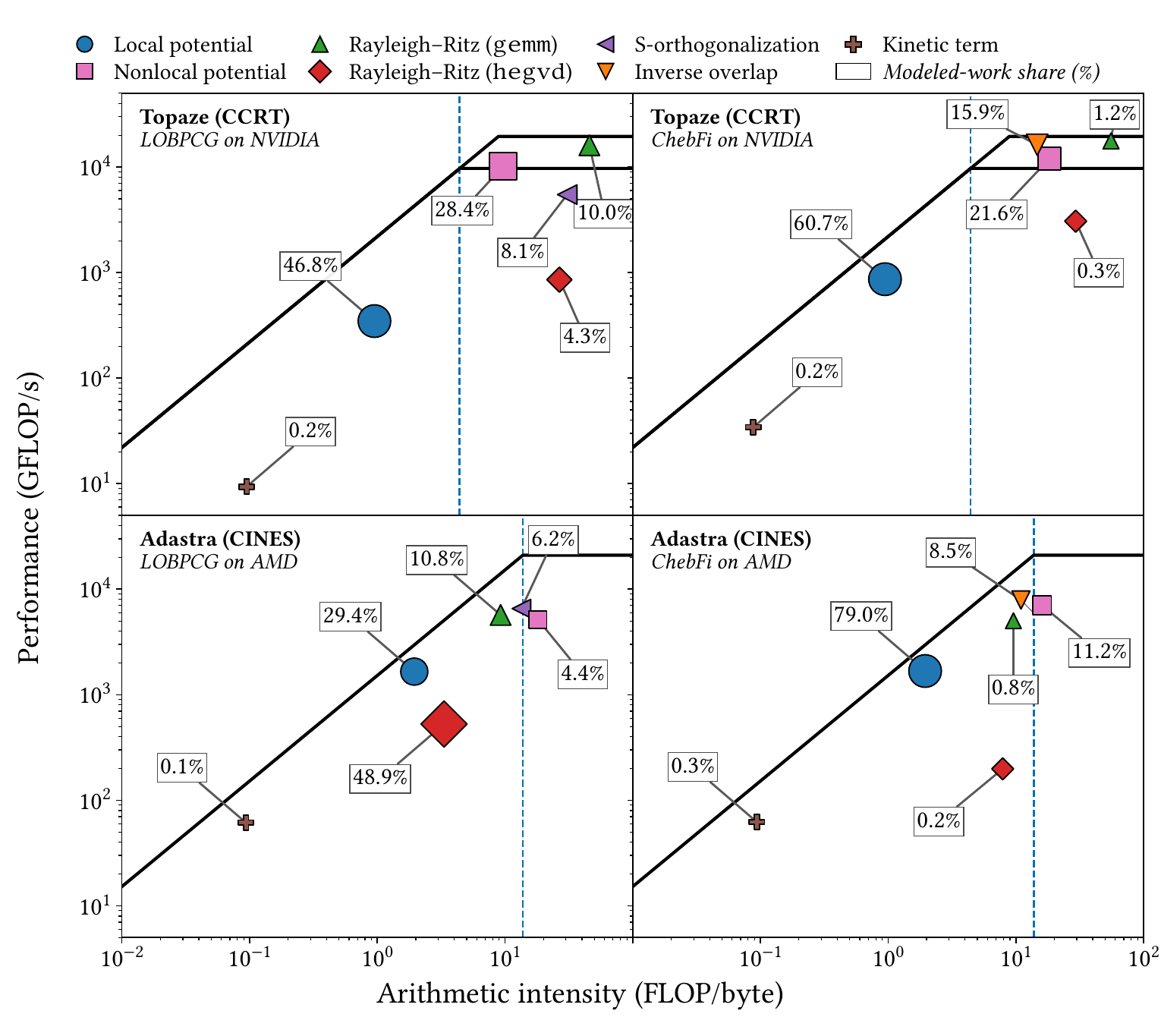}
    \caption{Roofline model for \lobpcg and \chebyshevfiltering (abbreviated ChebFi) algorithms at a single SCF step, using one \nvidia A100 GPU node (4 MPI, 4 GPU devices) and one \amd MI250X GPU node (8 MPI, 8 GPU devices). Dominant GPU kernels are plotted: (non-)local Hamiltonian-vector products ($V_{\text{(non)loc}}X$), kinetic-energy operator application, inverse overlap matrix-vector products ($S^{-1}X$), $S$-orthogonalization and the Rayleigh-Ritz procedure split into \gemm and diagonalization steps. The vertical blue dashed line separates the memory-bound region (left) from the compute-bound region (right). The two rooflines in \topaze represent the peak performance limits of \cuda cores and tensor cores, respectively. Kernel groups are ranked by their share of total rank-0 GPU-kernel work.}
    \label{fig:mpi_roofline}
\end{figure}

Figure~\ref{fig:mpi_roofline} presents the GPU roofline for kernel groups that contribute the most in the total FP64 performance and HBM memory traffic of the modeled GPU-kernel work. The percentage denotes their share of work.

The comparison between the two profilers should be interpreted qualitatively, as they collect hardware counters differently and the vendor-specific low-level libraries use different GPU kernel dispatch strategies. Moreover, although the aggregated Rayleigh-Ritz \texttt{hegvd} point lies in the compute-bound region on \topaze, its constituent kernel groups include both compute- and memory-bound operations. \texttt{hegvd} internally contains several memory-bound GPU kernels, primarily associated with tridiagonal reduction and transposition (\texttt{sytrd}/\texttt{latrd}), divide-and-conquer eigensolver steps (\texttt{laed}), Cholesky factorization (\texttt{potrf}), triangular solves (\texttt{trsm}), and matrix-vector updates (\texttt{gemv}/\texttt{gemvt}).

We distinguish several components of the Hamiltonian application and iterative diagonalization algorithms. The Hamiltonian application splits into a local part (using FFTs) and a non-local part (using \gemm operations). The non-local potential is compute-bound, providing evidence that the plane-wave Hamiltonian leverages BLAS3 operations. The kinetic-energy part of the Hamiltonian application is memory-bound as it mainly uses BLAS1 multiply-add operations (\texttt{axpy}). The Rayleigh-Ritz procedure is further decomposed to \gemm operations and to diagonalization in the trial eigenvector subspace.

The \gemm kernels in both algorithms leverage Tensor cores on \nvidia and are compute-bound, confirming their suitability for GPU acceleration. In contrast, the Rayleigh-Ritz solver, implemented through \hegvd calls, is a mixture of memory-bound and compute-bound internal linear algebra calls. The memory-bound parts dominate on \amd. The filtering step reaches peak performance of Tensor cores and is thus compute-bound, whereas several kernels fail to attain the memory roof due to data exceeding GPU capacity. To optimize overall performance, an algorithm should minimize both the number of calls and the space dimension of the Rayleigh-Ritz procedure.

From our roofline model benchmarks, the mathematical operations in iterative diagonalization algorithms fall into two categories~\cite{10.1145/1498765.1498785}:
\begin{itemize}
\item \textit{Compute-bound} operations: matrix-free Hamiltonian application and spectral polynomial filtering.
\item \textit{Memory-bound} operations: subspace orthogonalization and Rayleigh-Ritz procedure. These operate on relatively small matrices, having all bands and a part of plane waves. Due to their low arithmetic intensity, they do not scale and become a bottleneck.
\end{itemize}

\subsection{Single-step benchmarks: comparison of subspace iteration algorithms}\label{sec:per_gpu_si}

In the following, we examine how adjusting the accuracy of the iterative diagonalization algorithm, specifically by controlling the number of inner iterations, can substantially decrease the runtime. This can be achieved by tuning the block size in the \lobpcg method or the polynomial degree in \chebyshevfiltering. We show that, for the \chebyshevfiltering algorithm, increasing the polynomial degree improves the accuracy of the eigenvectors without degrading overall performance.

For the \lobpcg algorithm, the GPU speed-up at constant number of Hamiltonian applications is defined as $s = T^{\text{CPU}}_{\texttt{nline}} / T^{\text{GPU}}_{\texttt{nline}}$, where $T^{\text{CPU}}_{\texttt{nline}}$ (resp. $T^{\text{GPU}}_{\texttt{nline}}$) denotes the execution time on CPU (resp. GPU) for $\texttt{nline}$ minimization lines—i.e., $\texttt{nline}$ applications of the Hamiltonian operator per eigenvector.
The same indicator applies to the \chebyshevfiltering algorithm: $s = T^{\text{CPU}}_{\texttt{ndeg}} / T^{\text{GPU}}_{\texttt{ndeg}}$, where $T^{\text{CPU}}_{\texttt{ndeg}}$ (resp. $T^{\text{GPU}}_{\texttt{ndeg}}$) is the execution time on CPU (resp. GPU) when using a Chebyshev polynomial of degree $\texttt{ndeg}$.

\begin{figure}[H]
    \centering
    \includegraphics[width=0.6\linewidth]{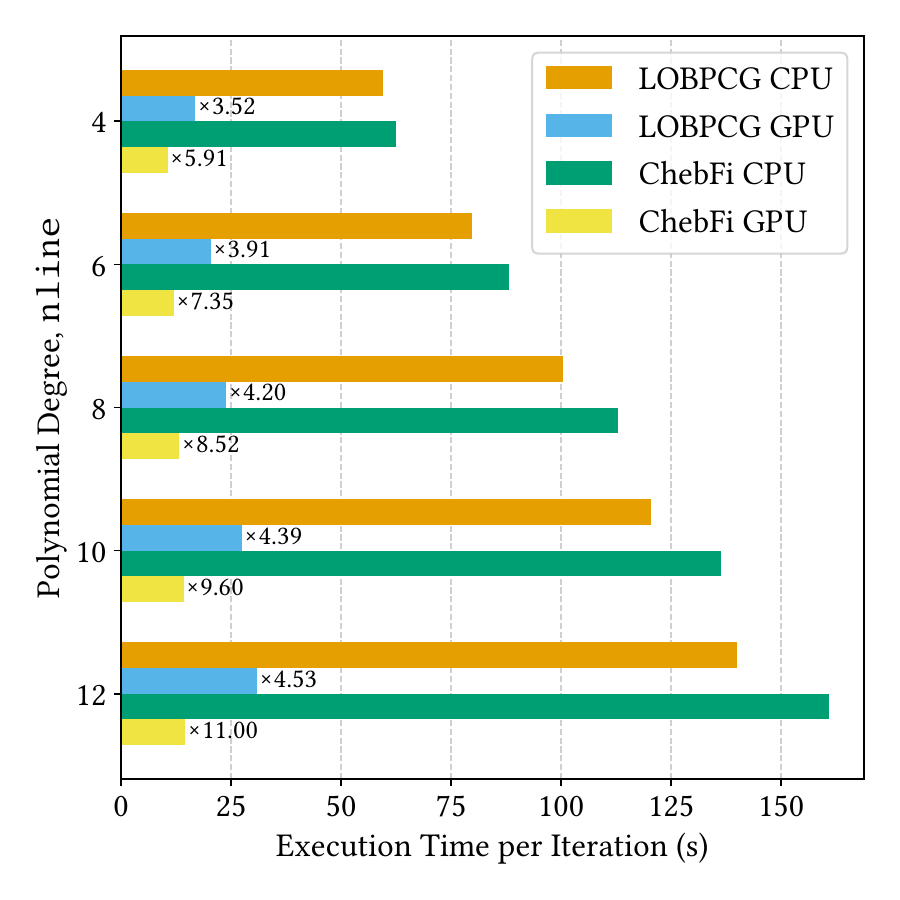}
    \caption{Execution times for a single SCF iteration on one CPU or GPU node, as a function of the number of minimization lines (\texttt{nline}) for \lobpcg or the polynomial degree (\texttt{ndeg}) for \chebyshevfiltering (abbreviated as ChebFi). The numbers next to the histograms indicate the CPU/GPU speedup at constant number of Hamiltonian applications, as defined in Section~\ref{sec:per_gpu_si}. These benchmarks were performed on the \topaze (CCRT) supercomputer using \amd Milan processors and \nvidia A100 GPUs (see \ref{sec:gpu_specifs}).}
    \label{fig:degree_periter}
\end{figure}

Figure~\ref{fig:degree_periter} shows the speedup of a single SCF iteration as accuracy increases, by comparing execution times on a single CPU or GPU node. Increasing the number of minimization lines in the \lobpcg algorithm has only a modest effect on speedup, whereas raising the polynomial degree in the \chebyshevfiltering algorithm significantly impacts performance. This indicates that \chebyshevfiltering incurs a lower execution-time penalty when increasing accuracy, compared to \lobpcg. Consequently, \chebyshevfiltering enables faster SCF convergence (fewer steps overall) at less performance cost, making it better suited to GPUs than \lobpcg.

Table~\ref{tab:chebfi_lobpcg_perf} shows that the accuracy gains from \chebyshevfiltering also reduce the final number of self-consistent field (SCF) iterations required for convergence.
In \lobpcg, the number of SCF iterations is already minimal even with few minimization lines, as the eigenvectors—particularly the last ones—are well converged. Thus, increasing the number of lines adds unnecessary computational workload without reducing SCF iterations further.
In contrast, raising the polynomial degree in \chebyshevfiltering substantially improves eigenvector convergence and thereby decreases the number of SCF iterations. Like \lobpcg, there is a minimal number of SCF iterations beyond which no further precision is gained, while computation time continues to rise with the polynomial degree. \chebyshevfiltering thus offers an optimal polynomial degree that maximizes GPU acceleration, though this critical value likely depends strongly on the simulated material.

\begin{table}[H]
\footnotesize
\centering
\begin{tabular}{lcccc}
\hline
Algorithm & Degree & Wall Time & \# SCF & Squared WF residual  \\ 
          & \texttt{ndeg} &   (s)        &  iterations & at iter. 1 \\ 
\hline
\multirow{6}{*}{ChebFi}
  & 4  & 295.8 & 28 &  3.0e-6 \\
  & 6  & 204.0 & 17 &  2.6e-6 \\
  & 8  & 198.8 & 15 &  1.5e-6 \\
  & 10 & 198.8 & 14 &  1.0e-6 \\
  & 12 & 204.7 & 14 &  8.0e-7 \\
  & 14 & 221.4 & 14 &  2.5e-7 \\
\hline
Algorithm & \texttt{nline} & Wall Time & \# SCF & Squared WF residual \\ 
          &  &     (s)      &  iterations & at iter. 1 \\ 
\hline
\multirow{5}{*}{\lobpcg}
  & 4  & 236.3 & \multirow{5}{*}{14} & 2.0e-7 \\
  & 6  & 285.6 &  &  3.5e-9 \\
  & 8  & 334.6 &  &  2.5e-10 \\
  & 10 & 384.4 &  &  2.5e-10 \\
  & 12 & 432.6 &  &  2.5e-10 \\
\hline
\end{tabular}
\caption{Execution times and number of SCF iterations in \abinit{} on a single GPU node, for ground-state determination of the energy, forces, and stress tensor in a 320-atom \ce{Ga2O3} crystal (1536 electronic bands). Results are shown as a function of the number of minimization lines (\texttt{nline}) for \lobpcg or the polynomial degree (\texttt{ndeg}) for \chebyshevfiltering (abbreviated as ChebFi). All runs used the same stopping criterion based on the density residual, with \lobpcg employing a fixed number of 4 blocks. Last column shows the maximum value of the squared wave function residual $|H\Psi_{nk} - \epsilon_{nk}S\Psi_{nk}|^2$ at the end of the first SCF iteration
These benchmarks were performed on the \topaze (CCRT) supercomputer using 4 \nvidia A100 GPUs (see \ref{sec:gpu_specifs}).}
\label{tab:chebfi_lobpcg_perf}
\end{table}

One major limitation of the \chebyshevfiltering algorithm on CPUs, as shown in Figure~\ref{fig:degree_periter}, is that the accuracy gains from increasing the polynomial degree fail to offset the filtering overhead per inner iteration. This leads to an overall slowdown in the SCF cycle execution time, despite fewer iterations.
In contrast, on GPUs, Table~\ref{tab:chebfi_lobpcg_perf} shows that raising the polynomial degree pays off due to highly accelerated Hamiltonian applications. The inner iterations then outperform even the best-case \lobpcg scenario, without significant accuracy loss. The last column of the table shows that LOBPCG achieves much higher accuracy at first inner iteration, without lowering the number of outer SCF iterations. Lastly, \chebyshevfiltering also provides finer control, offering more flexibility to balance accuracy and speed by gradually tuning the degree as needed.

\section{Conclusion and perspectives}

In this paper, we demonstrate that the GPU port of the \abinit software package enables highly efficient plane-wave density functional theory calculations. This GPU port was first made available in \abinit v10.0 and has been further improved and expanded in later versions. The complete GPU port we present in this paper is production-ready in \abinit v10.8 or later. Performance is excellent on both \nvidia GPUs and \amd MI-series GPUs, though it requires graphics accelerators with high double-precision computing capabilities. A simple porting effort alone does not ensure high performance; algorithmic adaptations are essential to make diagonalization procedures more GPU-optimized. Iterative diagonalization algorithms must be revisited to prioritize compute-bound operations with high arithmetic intensity, such as applying the Hamiltonian operator to large batches of wave functions. In this regard, spectrum polynomial filtering algorithms, such as \chebyshevfiltering, are the most suitable, as they build the eigenvector subspace through repeated Hamiltonian applications on extensive trial vector sets. In contrast, the \lobpcg algorithm, which relies on block-wise orthogonalization via parallel conjugate gradients, remains more memory-bound.

Memory-bound operations also limit multi-GPU scaling. For this reason, the Rayleigh-Ritz procedure should be performed as rarely as possible, on the smallest eigenvector subspace. Theoretical cost models and \abinit benchmarks both confirm that matrix-matrix multiplications on large wave-function sets boost arithmetic intensity and should thus be favored. We notice that the systems studied in this work are currently limited mainly by GPU memory. This limitation is expected to be less restrictive in the future by GPUs with larger memory.

To reduce the impact of compute-bound subspace diagonalization, a viable solution proposed in the literature is the Spectrum Slicing method \cite{LIOU2020107330,SCHOFIELD2012497}. This approach enables fast eigensolutions for large batches of Hermitian matrices processed in parallel on GPUs, while substantially reducing the size of the Rayleigh-Ritz operation. This method would facilitates simulations of larger systems by distributing wave functions across GPUs during Rayleigh-Ritz steps. It avoids a global Rayleigh-Ritz procedure across all bands, instead computing smaller independent eigenvalue slices. In such polynomial filter-based methods, the polynomial degree must be much higher than in standard \chebyshevfiltering. As demonstrated in this paper, the numerous Hamiltonian applications involved likely have only limited overhead on GPUs.

\section*{Acknowledgments}

This project was provided with HPC and storage resources by GENCI at CINES and IDRIS computing centers thanks to the grant 2023-AD010914563 on the supercomputers \jeanzay and \adastra. Part of the work was performed using HPC resources from CCRT (CEA, France) on the \topaze supercomputer.
This project was partly supported by the European Union’s Horizon 2020 research and innovation program under the grant agreement No 951786 (NOMAD CoE).
The authors are grateful to Cl\'ementine Barat for insightful discussions regarding \chebyshevfiltering algorithm.

\appendix

\section{Machine specifications}
\label{sec:gpu_specifs}

Technical specifications of the supercomputers used for benchmarking are summarized in Table~\ref{tab:gpu_specifs}.

\begin{table}[H]
    \footnotesize
    \centering
    \begin{tabular}{l|c|c}
    \toprule
    \textbf{Supercomputer} & \topaze & \adastra \\ \midrule
    Center                 & CCRT (France)~\cite{CCRT2026} & CINES (France)~\cite{CINES2026} \\ 
    Model                  & BullSequana XH2000 & HPE Cray EX4000 \\

    \midrule
    \textbf{CPU-only partition} & \textit{Milan} & \textit{Genoa} \\ \midrule
  
    \# Nodes                   & 996 & 544 \\
    Processors per node        & 2 \amd EPYC Milan 7763 & 2 \amd EPYC Genoa 9654 \\
    CPU frequency              & 2.45 GHz & 2.40 GHz \\
    \# CPU cores per node      & 128 & 192 \\
    RAM per node               & 256 GB & 768 GB \\
    Peak performance per node  & 4.72 TFlop/s & 7.37 TFlop/s \\
    Network model              & InfiniBand HDR-100 & Slingshot-11 \\
    Network bandwidth per node & 100 Gbit/s & 200 Gbit/s \\

    \midrule
    \textbf{CPU-GPU partition} & \textit{A100} & \textit{MI250X} \\ \midrule

    \# Nodes                            & 75 & 356 \\
    Host processor per node             & 2 \amd EPYC Milan 7763 & 1 \amd EPYC Trento 7A53 \\
    CPU frequency                       & 2.45 GHz & 2.00 GHz \\
    \# CPU cores per node               & 128 & 64 \\
    GPUs per node                       & 4 \nvidia A100 SXM4 & 4 \amd Instinct MI250X \\
    Memory per GPU                      & 80 GB & 128 GB \\
    GPU memory per node        & 320 GB & 512 GB  \\
    GPU memory bandwidth                & 2 TB/s & 3.2 TB/s \\
    Peak performance per node  & 57.33 TFlop/s & 193.54 TFlop/s \\
    Network model                       & InfiniBand HDR & Slingshot-11 \\
    Network bandwidth per node & 200 Gbit/s & 800 Gbit/s \\

\bottomrule
\end{tabular}
    \caption{Technical specifications of CPU-only and CPU-GPU partitions of supercomputers used in the present work. Peak performance is for FP64 operations.}
    \label{tab:gpu_specifs}
\end{table}
  
\section{\abinit GPU-enabled functionalities} \label{sec:gpu_features}

Table~\ref{tab:gpu_features} summarizes GPU-enabled functionalities in \abinit version 10.8. As a reference, Table~\ref{tab:gpu_features} also provides representative performance figures for selected ``beyond-DFT'' calculations in \abinit on GPUs. For ``beyond-DFT'' methods, batching opportunities are more limited than in standard ground-state DFT, resulting in lower arithmetic intensity and, consequently, reduced GPU performance. The reported average speedups correspond to the indicated GPU models and depend on the hardware configuration, memory setup, and problem size. The \nvidia (respectively, \amd) results were obtained on the \topaze (respectively, \adastra) supercomputer. These data should be regarded as illustrative rather than exhaustive. Typical test cases: 31-atom Au system for hybrid-functional and DFPT calculations, and 54-atom Fe system for DMFT calculations.

\begin{table}[H]
\centering
\footnotesize
\begin{tabular}{>{\raggedright\arraybackslash}p{0.3\linewidth}|l|cc}
    \toprule
    \multirow{2}{*}{\centering Topic}
    & \multirow{2}{*}{Property or method}
    & \multicolumn{2}{l}{\shortstack{Typical CPU/GPU\\speedup range}} \\
    \cline{3-4}
    & & \nvidia & \amd \\
    \midrule

    \multirow{5}{=}{\raggedright Ground-state properties\newline (DFT)}
    & Total energy                         & \multirow{5}{*}{8--12} & \multirow{5}{*}{5--7}  \\
    & Forces, stress tensor                & & \\
    & \lobpcg algorithm                    & & \\
    & \chebyshevfiltering algorithm        & & \\
    & DFT+U                                & & \\
    \midrule

    Beyond DFT properties & Hybrid functionals & 8--10 & 4--8 \\
    \midrule

    \multirow{6}{=}{\raggedright Response functions\newline (DFPT)}
    & Response to perturbation of \kpoint  & \multirow{6}{*}{1.5--3} & \multirow{6}{*}{1--2} \\
    & Response to perturbation of electric field
                                         & & \\
    & Atomic vibrations (phonons at $q=0$) & & \\
    & Atomic vibrations (phonons at $q\neq0$) & & \\
    & Elastic tensor                       & & \\
    & Metals                               & & \\
    \midrule

    \multirow{3}{=}{\raggedright
    Many-body theory\newline (DMFT)}
    & Green function computation           & \multirow{3}{*}{3--6} & \multirow{3}{*}{5--8} \\
    & Hubbard-I solver                     & & \\
    & CT-QMC solver (partially ported)     & & \\
    \bottomrule
\end{tabular}
\caption{\abinit functionalities ported to GPUs as of version 10.8, grouped by physical property, for DFT and ``beyond-DFT'' calculations. All listed features support norm-conserving pseudopotentials, the projector-augmented-wave (PAW) method, collinear and non-collinear magnetism, and spin--orbit coupling.}
\label{tab:gpu_features}
\end{table}

\section{Parallel subspace iteration algorithms}\label{sec:parallel_algos_detail}

We provide the detailed parallel \lobpcg and \chebyshevfiltering algorithms as implemented in \abinit. Every step including MPI communications is indicated (all-to-all \mpia, reduction \mpir or small reduction \mpisr). Algorithm~\ref{alg:lobpcg_k} shows parallel matrix-free \lobpcg algorithm in \abinit, as in \cite{BOTTIN2008329}. In the line minimization loop, $K$ denotes the preconditioner (diagonal matrix) and $P$ is a matrix of conjugate vectors to $\Psi$ and $W$ computed in the Rayleigh-Ritz procedures \cite{BOTTIN2008329,knyazev2001,PETSc}.  $\Lambda$ is a diagonal matrix with Ritz values in its diagonal. Algorithm~\ref{alg:chebfi_k} shows parallel matrix-free \chebyshevfiltering.
    
\begin{algorithm}[H]
    \centering
    \footnotesize
    \setlength{\arrayrulewidth}{0.3pt} 
    \setlength{\tabcolsep}{4pt}        
    \begin{tabular}{|l|}
    \hline
    \begin{minipage}[t]{0.9\linewidth}
    \begin{algorithmic}[1]
        \Require{$k$ line minimizations, $M$ starting linearly independent electronic bands in $\Psi$ partitioned into $b$ blocks.}
        \Ensure{$\Lambda,\Psi$ solution to $H\Psi = \Lambda S\Psi$, residual error norm.}
        \For{$\Psi_i \in \Psi_1,\ldots,\Psi_b$}
        \If{$i>1$}
            \State \mpir $\Psi_i\gets\sdeflate(n_p,M/b,(i-1)M/b)$
        \EndIf
        \State \mpia Transpose Row-to-Column.
        \State $\ttH\Psi_i,\ttS\Psi_i \gets \texttt{HX\_SX}(N,m_p/b)$
        \State \mpia Transpose Column-to-Row.
        \State \mpir $\Psi_i,\ttH\Psi_i,\ttS\Psi_i\gets \sortho(n_p,M/b)$
        \State \mpir $\Lambda_i,\Psi_i,\ttH\Psi_i,\ttS\Psi_i\gets \rr(n_p,M/b)$
        \For{$n=1,\ldots,k$}
            \State $W_i \gets (H-\Lambda_iS)\Psi_i$
            \State \mpisr Compute residual norms and exit if block is converged.
            \State $W_i \gets KW_i$
            \If{$i>1$}
                \State \mpir $W_i\gets\sdeflate(n_p,M/b,(i-1)M/b)$
            \EndIf
            \State \mpia Transpose Row-to-Column.
            \State $\ttH W_i,\ttS W_i \gets \texttt{HX\_SX}(N,m_p/b)$
            \State \mpia Transpose Column-to-Row.
            \If{$n=1$}
                \State \mpir $\Psi_i, \ttH\Psi_i, \ttS\Psi_i, W_i, \ttH W_i, \ttS W_i \gets \sortho(n_p,2M/b)$
                \State \mpir $\Lambda_i,\Psi_i, \ttH\Psi_i, \ttS\Psi_i, P_i, \ttH P_i, \ttS P_i \gets \rr(n_p,2M/b)$
            \Else
                \State \mpir $\Psi_i, \ttH\Psi_i, \ttS\Psi_i, W_i, \ttH W_i, \ttS W_i, P_i, \ttH P_i, \ttS P_i \gets \sortho(n_p,3M/b)$
                \State \mpir $\Lambda_i,\Psi_i,\ttH\Psi_i, \ttS\Psi_i, P_i, \ttH P_i, \ttS P_i \gets \rr(n_p,3M/b)$
            \EndIf
            \State $\Psi_i \gets \Psi_i + P_i$, $\ttH\Psi_i \gets \ttH\Psi_i + \ttH P_i$, $\ttS\Psi_i \gets \ttS\Psi_i + \ttS P_i$
        \EndFor
        \State \mpisr Compute residual norms.
        \EndFor
        \If{$b>1$}
        \State \mpir $\Psi,\ttH\Psi,\ttS\Psi \gets \sortho(n_p,M)$
        \State \mpir $\Lambda,\Psi,\ttH\Psi,\ttS\Psi\gets\rr(n_p,M)$
        \EndIf
    \end{algorithmic}
    \InnerAlgorithmCaption{Parallel \lobpcg with $b$ blocks and $k$ line minimizations.}{alg:lobpcg_k}
    \end{minipage}
    \\ \hline
  \end{tabular}
\end{algorithm}

\begin{algorithm}[H]
    \centering
    \footnotesize
    \setlength{\arrayrulewidth}{0.3pt} 
    \setlength{\tabcolsep}{4pt}        
  
    \begin{tabular}{|l|}
    \hline
    \begin{minipage}[t]{0.9\linewidth}
    \begin{algorithmic}[1]
        \Require{$k$ polynomial degree, $M$ starting linearly independent electronic bands in $\Psi$.}
        \Ensure{$\Lambda,\Psi$ solution to $H\Psi = \Lambda S\Psi$, residual error norm.}
        \State \mpia Transpose Row-to-Column.
        \State $\ttH\Psi,\ttS\Psi \gets \texttt{HX\_SX}(N,m_p)$
        \State \mpisr Compute Rayleigh quotients and get minimum and maximum.
        \For{$n=1,\ldots,k$}
            \State $\ttS^{-1}\ttH\Psi\gets\texttt{Sinv}(N,m_p)$
            \State $\Psi \gets \texttt{axpy}(N,m_p)$
            \State $\ttH\Psi,\ttS\Psi \gets \texttt{HX\_SX}(N,m_p)$
        \EndFor
        \State \mpia Transpose Column-to-Row.
        \State \mpir $\Lambda,\Psi,\ttH\Psi,\ttS\Psi\gets\rr(n_p,M)$
        \State $W \gets (H-\Lambda S)\Psi$
        \State \mpisr Compute residual norms.
    \end{algorithmic}
    \InnerAlgorithmCaption{Parallel \chebyshevfiltering of degree $k$.}{alg:chebfi_k}
    \end{minipage}
    \\ \hline
    \end{tabular}
\end{algorithm}

\bibliography{cas-refs}
\bibliographystyle{abbrv}

\end{document}